\documentclass[useAMS,usedcolumn,usenatbib,usegraphicx]{mn2e}

\def\leff{\ifmmode \lambda_{\mathrm{eff}}\else $\lambda_{\mathrm{eff}}$\fi}
\def\oii{\ifmmode \rmn{[O}\,\textsc{ii}\rmn{]} \else [O\,\textsc{ii}]\fi}

\def\ha{\ifmmode \rmn{H}{\alpha}\else H$\alpha$\fi}
\def\hb{\ifmmode \rmn{H}{\beta}\else H$\beta$\fi}
\def\ebmv{\ifmmode E(B-V) \else $E(B-V)$\fi}
\def\ebmvg{\ifmmode E(B-V)_{\rmn{g}} \else $E(B-V)_{\rmn{g}}$\fi}
\def\ebmvs{\ifmmode E(B-V)_{\rmn{s}} \else $E(B-V)_{\rmn{s}}$\fi}
\def\ergsa{\ifmmode \rmn{erg}\,\rmn{s}^{-1}\,\mbox{\AA}^{-1}\else $\rmn{erg}\,\rmn{s}^{-1}\,\mbox{\AA}^{-1}$\fi}
\def\ergshz{\ifmmode \rmn{erg}\,\rmn{s}^{-1}\,\rmn{Hz}^{-1}\else $\rmn{erg}\,\rmn{s}^{-1}\,\rmn{Hz}^{-1}$ \fi}
\def\ergscma{\ifmmode \rmn{erg}\,\rmn{s}^{-1}\,\rmn{cm}^{-2}\,\mbox{\AA}^{-1}\else $\rmn{erg}\,\rmn{s}^{-1}\,\rmn{cm}^{-2}\,\mbox{\AA}^{-1}$\fi}

\def\ergscmhz{\ifmmode \rmn{erg}\,\rmn{s}^{-1}\,\rmn{cm}^{-2}\,\rmn{Hz}^{-1}\else $\rmn{erg}\,\rmn{s}^{-1}\,\rmn{cm}^{-2}\,\rmn{Hz}^{-1}$\fi}
\def\ergs{\ifmmode \mbox{erg}\,\mbox{s}^{-1}\else $\mbox{erg}\,\mbox{s}^{-1}$\fi}
\def\pegase2{\textsc{pegase-ii}}
\def\mperyr{\ifmmode \rmn{M}_{\odot}\,\rmn{yr}^{-1}\else $\rmn{M}_{\odot}\,\rmn{yr}^{-1}$\fi}
\def\msun{\ifmmode \rmn{M}_{\odot}\else $\rmn{M}_{\odot}$\fi}
\def\lha{\ifmmode \rmn{L}_{\ha} \else L$_{\ha}$\fi}
\def\luv{\ifmmode \rmn{L}_{\rmn{UV}} \else L$_{\rmn{UV}}$\fi}
\def\lu{\ifmmode \rmn{L}_{u'} \else L$_{u'}$\fi}

\newcommand{\iraf}{\textsc{iraf}}

\newcommand{\sex}{\textsc{sextractor}}
\newcommand\om{\ifmmode \Omega_{\rmn{M}}\else $\Omega_{\rmn{M}}$\fi}
\newcommand\ok{\ifmmode \Omega_{\rmn{k}}\else $\Omega_{\rmn{k}}$\fi}
\newcommand\ol{\ifmmode \Omega_{\Lambda}\else $\Omega_{\Lambda}$\fi}
\def\aj{AJ}                   
\def\araa{ARA\&A}             
\def\apj{ApJ}                 
\def\apjl{ApJ}                
\def\apjs{ApJS}               
\def\aap{A\&A}                
\def\aaps{A\&AS}              
\def\mnras{MNRAS}             
\def\pasp{PASP}               
\def\procspie{Proc.~SPIE}   

\title[An ultraviolet-selected galaxy redshift survey - III]{An ultraviolet-selected galaxy redshift survey - III: Multicolour imaging and non-uniform star formation histories} 
\author[M. Sullivan et al.]
{Mark~Sullivan$^{1,2}$\thanks{E-mail: sullivan@astro.utoronto.ca},
  Marie~A.~Treyer$^{3}$,
  Richard~S.~Ellis$^4$,
  Bahram Mobasher$^5$\\
  $^1$ Department of Physics, University of Durham, South Road, Durham, DH1 3LE, UK\\
  $^2$ Department of Astronomy and Astrophysics, University of Toronto, 60 St. George Street, Toronto, ON M5S 3H8, Canada\\
  $^3$ Laboratoire d'Astrophysique de Marseille, Traverse du Siphon, 13376 Marseille, France\\
  $^4$ California Institute of Technology, E. California Blvd, Pasadena, CA 91125,
  USA\\
  $^5$ Space Telescope Science Institute, 3700 San Martin Drive, Baltimore, MD 21218, USA\\
}
\date {Accepted ---. Received ---; in original form ---.}

\begin{document}
\label{firstpage}
\maketitle

\begin{abstract}
  We present panoramic $u$' and optical ground-based imaging
  observations of a complete sample of low-redshift ($0<z<0.4$)
  galaxies selected in the ultraviolet (UV) at 2000\,\AA\ using the
  balloon-borne FOCA instrument of Milliard et al. This survey is
  highly sensitive to newly-formed massive stars, and hence to
  actively star-forming galaxies.  We use the new data to further
  investigate the optical, stellar population and star formation
  properties of this unique sample, deriving accurate galaxy types and
  $k$-corrections based on the broad-band spectral energy
  distributions.
  
  When combined with our earlier spectroscopic surveys, these new data
  allow us to compare star-formation measures derived from
  aperture-corrected \ha\ line fluxes, UV(2000\,\AA) and
  $u$'(3600\,\AA) continuum fluxes on a galaxy-by-galaxy basis.  As
  expected from our earlier studies, we find broad correlations over
  several decades in luminosity between the different dust-corrected
  star-formation diagnostics, though the scatter is larger than that
  from observational errors, with significant offsets from trends
  expected according to simple models of the star-formation histories
  (SFHs) of galaxies.  Popular galaxy spectral synthesis models with
  varying metallicities and/or initial mass functions seem unable to
  explain the observed discrepancies.
  
  We investigate the star-formation properties further by modelling
  the observed spectroscopic and photometric properties of the
  galaxies in our survey. We find that nearly half of the galaxies
  surveyed possess features that appear incompatible with simple
  constant or smoothly declining SFHs, favouring instead irregular or
  temporally-varying SFHs.  We demonstrate how this can reconcile the
  majority of our observations, enabling us to determine empirical
  corrections that can be used to calculate intrinsic star formation
  rates (as derived from \ha\ luminosities) from measures based on UV
  (or $u$') continuum observations alone. We discuss the broader
  implications of our finding that a significant fraction of
  star-forming galaxies have complex SFHs, particularly in the context
  of recent determinations of the cosmic SFH.

\end{abstract}


\section{Introduction}
\label{sec:introduction}

An accurate determination of the star-formation history (SFH) of the
Universe is one of the key goals of modern observational cosmology.
Studies which measure, analyse and model its precise form are
important not only in indicating likely epochs of dominant activity,
but also for comparisons with the predictions of semi-analytical
models of galaxy formation, and as such have provided a major impetus
towards a fuller understanding of the physical mechanisms of galaxy
evolution.

Various techniques now exist to measure star-formation rates (SFRs) in
distant galaxies, all of which are sensitive in some way to the number
of young, short-lived, and hence massive stars. These include nebular
recombination (e.g. \ha) and forbidden line (e.g. \oii) emission
\citep{1983ApJ...272...54K,1989AJ.....97..700G,1995ApJ...455L...1G,2002MNRAS.337..369T,2003A&A...402...65H},
ultraviolet (UV) continuum measures in the wavelength range 1500 to
2800\,\AA\ 
\citep{1987A&A...180...12D,1996ApJ...460L...1L,1997ApJ...486L..11C,1998MNRAS.300..303T,1999AJ....118..603C,2000MNRAS.312..442S,2002AJ....124.1258W},
decimetric radio emission
\citep{1992ARA&A..30..575C,1998ApJ...507..155C,2000ApJ...544..641H},
and far-infrared continuum emission
\citep{1997MNRAS.289..490R,1999MNRAS.302..632B,2002AJ....124.3135K}
\citep[for a review of these techniques and their associated
uncertainties, see][]{1998ARA&A..36..189K}. Currently, many of these
diagnostics are only available over limited (and sometimes
non-overlapping) redshift ranges, so studies of the evolution of the
cosmic SFH have tended to combine many disparate studies measuring
star-formation using a selection of the techniques described above,
without an accurate understanding of how underlying physical processes
in galaxies can affect different diagnostics.

However, there has been recent progress in inter-comparing and
calibrating these measures by investigating the degree to which the
different diagnostics agree when measured for the same galaxies. Work
to date comparing \ha/UV measures
\citep{1999MNRAS.306..843G,2000MNRAS.312..442S,2001AJ....122..288H,2001ApJ...548..681B,2002A&A...383..801B}
and radio/\ha/UV measures
\citep{1998ApJ...507..155C,2001AJ....122..288H,2001ApJ...558...72S}
has revealed broad correlations but with a significant scatter,
together with some offsets from relations expected from simple SFH
scenarios. Such discrepancies indicate either poorly understood dust
extinction corrections (particularly on UV measures), some physical
parameter varying from galaxy to galaxy, such as the metallicity or
initial mass function, or variations in the timescale of recent
star-formation and consequent contamination of UV-continuum derived
SFRs by older and less massive stars.

We have been using a unique sample of local UV-selected galaxies to
investigate these issues, constructed using the balloon-borne camera of
\citet{1992A&A...257...24M} which images the sky at 2000\,\AA. In
\citet{1998MNRAS.300..303T} we used this sample to construct the first
local UV luminosity function, finding an integrated star-formation
density higher than that found in local emission-line surveys.
\citet{2000MNRAS.312..442S} extended this sample, and performed
initial investigations into the physical nature of star-formation in
the sample comparing \ha\ and UV-derived SFRs. Evidence for
non-linearities and significant scatter was found, though questions
over aperture corrections on the \ha\ measures and $k$-corrections on
the UV luminosities remained. The radio properties
\citep{2001ApJ...558...72S} and chemical properties
\citep{2002MNRAS.330...75C} of this sample have also been
investigated.

In this paper, we further constrain the nature of star-formation in
the sample. We extend our previous studies to include new broad-band
$u$' and optical $Br'$ imaging observations of a substantial fraction
of the galaxies in the survey, allowing accurate aperture and
$k$-corrections to be made. We then use the combined samples to
investigate the optical and stellar population properties of the
sample, as well as to investigate SFRs derived from a second UV
continuum measure in conjunction with that previously measured at
2000\,\AA.

An outline of the paper is as follows. In Section~\ref{sec:datasets}
we introduce the redshift survey data, including the new imaging
campaigns. Section~\ref{sec:optical-properties} discusses the optical
properties of the sample, allowing improved aperture and
$k$-corrections for our spectroscopic data.
Section~\ref{sec:star-form-prop} then examines the star-formation
properties of the galaxies derived from \ha/UV/$u$' observations. We
model and discuss the implications of our results in
Section~\ref{sec:modelling}, and conclude in
Section~\ref{sec:conclusions}. Throughout this paper we assume a
$\ol=0.7$, $\om=0.3$, $h=0.70$ (where
$H_0=100\,h\,\rmn{km\,s^{-1}\,Mpc}^{-1}$) cosmology.

\section{The Survey Datasets}
\label{sec:datasets}

This paper extends our earlier work of \citet[][hereafter
T98]{1998MNRAS.300..303T} and \citet[][hereafter
S2000]{2000MNRAS.312..442S} by augmenting our photometric dataset with
with more extensive and precise optical photometry.  Here, we
introduce the new datasets which are analysed in this paper.  We
commence with a brief discussion of the existing datasets -- UV
imaging data taken from the FOCA experiment and the follow-up optical
spectroscopy -- followed by the details of the new imaging surveys of
these fields.

\subsection{The FOCA redshift survey}
\label{sec:existing-datasets}

The FOCA instrument is a balloon-borne 40-cm Cassegrain with a single
filter approximating a Gaussian centred at 2015\AA, FWHM 188\AA\ 
\citep{1992A&A...257...24M}. Two fields observed by FOCA are studied
here: the high galactic-latitude region of Selected Area 57 (SA57;
$\rmn{RA}=13^{\rmn{h}} 06^{\rmn{m}} 10\fs43$, $\rmn{Dec.}=+29\degr
01\arcmin 46\farcs1$, J2000) and the field of the $z=0.0215$ Leo
cluster, Abell 1367 (A1367; $\rmn{RA}=11^{\rmn{h}} 45^{\rmn{m}}
20\fs92$, $\rmn{Dec.}=+19\degr 53\arcmin 23\farcs0$, J2000).  The UV
exposure times corresponded to a limiting magnitude of $m_{\rmn
  UV}=18.5$ in the FOCA photometric system ($m_{\rmn UV}^{\rmn
  AB}\simeq20.75$).  \footnote{Subsequent to the publication of S2000,
  the FOCA instrument team have re-calibrated the conversion required
  to transform FOCA instrumental magnitudes to an external absolute
  magnitude system, resulting in a minor adjustment in the sense that
  $m_{\rmn{new}}=m_{\rmn{s2000}}+0.02$.}  Both fields have extensive
multi-fibre optical spectroscopy obtained using the WIYN/Hydra
(3500\,--\,6600\,\AA; 3.1-arcsec diameter fibres) and WHT/WYFFOS
(3500\,--\,9000\,\AA; 2.7-arcsec diameter fibres) telescope/instrument
combinations, providing redshifts for 224 UV-selected emission-line
galaxies, of which \ha\ fluxes can be reliably measured in 111
objects. The wavelength coverage of the WIYN data preclude observation
of \ha. Fluxes, equivalent widths (EWs), and errors for each of the
principle emission lines, as well as the strength of the 4000\,\AA\ 
Balmer break \citep[D4000; see e.g.][]{1983ApJ...273..105B}, are
measured where the signal-to-noise (S/N) of the spectra permits. Where
\ha\ is not detected, an upper flux limit is estimated at the emission
wavelength using the noise characteristics of each spectrum.

The Balmer emission lines are corrected for the effects of stellar
absorption as described in earlier papers (S2000), with a median
correction of 2.5\,\AA.  Fluxes are also corrected for the effects of
Galactic extinction using the dust maps of \citet{1998ApJ...500..525S}
and a \citet{1989ApJ...345..245C} extinction law. These corrections
are small, with typical \ebmv s of 0.01 and 0.02 for SA57 and A1367
respectively. The ratio of these corrected \ha\ and \hb\ fluxes is
used to determine the colour excess of the ionised gas, \ebmvg, and
hence the internal dust extinction assuming case-B recombination and a
\citet{1989ApJ...345..245C} attenuation law. AGN and QSO-like objects
are removed from the star-forming sample on the basis of their optical
spectra (objects with broad emission lines are discarded). Of the
remaining narrow-band objects, the vast majority are star-forming
galaxies, as evidenced from diagnostic diagrams based upon the ratios
of their principle emission lines \citep{2002MNRAS.330...75C}. Details
of all these procedures can be found in \citet{2002Obs...122..307S}.

\subsection{Optical Imaging Campaigns}
\label{sec:palom-imag-camp}

The new optical data are taken from three sources: A $u$'/$r$'
Palomar/LFC imaging programme of SA57, $B$-band imaging using the
CFHT/12k camera, and supplementary $gri$ data from the Palomar Digital
Sky Survey (DPOSS) for both SA57 and A1367.

\subsubsection{The Palomar Large Format Camera}
\label{sec:palomar-large-format}

The Large Format Camera \citep[LFC][]{2000AAS...196.5209S} on the
Palomar 200-in Hale telescope is a mosaic of six 2048x4096 pixel CCDs
(though only four were available when the data for this paper were
collected), covering a region of diameter $\sim24$$\arcmin$ with a
pixel scale of 0.175 arcseconds.  Data were collected over the course
of six dark nights split over two observing runs in April 2000 and
March 2001. The 1.5$\degr$ diameter field of SA57 was mosaiced using
Sloan Digital Sky Survey (SDSS) $u$' and $r$' filters, with a
tessellation pattern avoiding bright foreground stars which created
serious saturation and bleed trails on the LFC CCDs. Exposure times
were 2400s in $u$' and 600s in $r$'. Due to time constraints, we
concentrated our observations on the centre of the SA57 field.

The LFC data were reduced using a hybrid of the National Optical
Astronomy Observatories (NOAO) mosaic reduction software for use in
\iraf\ \citep{1998adass...7...53V} as well as custom-written routines
to deal with some issues specific to the LFC. The data were
bias-subtracted using the overscan regions on the chips and nightly
master bias frames.  Flat-fielding was performed using dome flats (for
$r$' data) and twilight sky flats (for $u$' data) taken on each night.
We also constructed a `super-sky' flat for the $r$' data by stacking
all the $r$' observations; however the resulting sky flat-field frame
was virtually indistinguishable from the dome flat, and hence was not
used. The \iraf\ routine \textsc{mscskysub} was used to remove any
second-order sky gradients across the field by computing medians in
boxes of size 50 pixels and fitting a low-order two dimensional
function to the median points.  The residual `fitted-sky minus sky
mean' is subtracted from each pixel.

We derived astrometric solutions for each chip in the two filters
using astrometric calibration exposures.  SA57 is a (deliberately
chosen) high galactic latitude field where the surface density of
bright foreground galactic stars required for astrometric calibration
is low.  We therefore observed astrometric calibration fields taken
from the Astrometric Calibration Regions (ACRs) catalogue of
\citet{1999AJ....118.2488S}, with star positions typically accurate to
$\pm26$\,mas.  We use \sex\ \citep{1996A&AS..117..393B} and the
WCSTools software suite written by D. Mink
\citep[e.g.][]{1999adass...8..498M} to create approximate world
co-ordinate system (WCS) solutions for each astrometric region in each
filter based on stars in the ACR catalogues, discarding stars with
large proper motions.  The WCS was then refined using the \iraf\ 
program \textsc{ccmap} and a TNX projection with a high order
polynomial and half cross-terms. This flexible function was needed to
fit the edges of the LFC chips where distortion can be significant.
The plate solution for each filter/chip was then applied to every
science exposure using the \iraf\ \textsc{mscred} tasks, shifting the
reference point for the plate solution for each science field using
the USNO-A2.0 stars (i.e. there are enough of these stars to calculate
a reference point shift, but not to derive a full plate solution). The
final solution for each science field is accurate to
0.15\,--\,0.2\arcsec, even for the regions of the chips that suffer
significant distortion.

Each exposure was projected and re-sampled onto a linear WCS, and the
individual dither steps at each pointing median-combined after masking
cosmic-rays and chip defects to create one final dithered image for
each chip, applying multiplicative scaling factors to allow for
airmass differences between dithers. The result is an astrometrically
calibrated and photometrically stable combined science image.

At the time the observing program was carried out, the SDSS filters
were `non-standard', in the sense that few published calibration stars
were available.  To avoid complex colour terms needed to place our
data on the standard $\alpha$-Lyrae photometric system, we used
standard spectrophotometric stars to calibrate our data. On each
night, two spectrophotometric standard stars were monitored at
approximately two hour periods throughout the night to derive
airmass-dependent extinction correction terms, augmented by dawn and
dusk twilight observations of 4\,--\,5 other spectrophotometric
standard stars. Our preferred standards were Feige 67 and Feige 34. We
calculated the AB magnitude of each standard star in the $u$' and $r$'
filters using the spectrophotometric star spectra energy distributions
(SEDs) taken from \citet{1988ApJ...328..315M} and
\citet{1990ApJ...358..344M}. We convolved these SEDs with the filter
and LFC-CCD responses (C. Steidel, private communication) to calculate
the magnitude in a particular filter.

The $r$' zero-point was extremely stable with a mean r.m.s. dispersion
of $\simeq0.01$\,mag. The $u$' zero-point was less well-defined, with
an r.m.s.  variation from standard star to standard star of
$\simeq0.03$\,mag. Due to the accuracy limit to which the spectra of
these standard stars are calibrated an accuracy greater than 0.03\,mag
in $u$' is probably not achievable \citep[see
e.g.][]{1993AJ....105.2017S}. As an additional consistency check, the
overlapping nature of the mosaicing technique allows us to compare
magnitudes of objects in different fields, and a night-to-night (or
even year-to-year) comparison is possible. We typically find agreement
to within 0.02\,mag in $r$', rising to 0.04\,mag for the $u$' data. No
offsets were seen between the same objects measured on different
nights.

Object catalogues were created using \sex\ version 2.2.2
\citep{1996A&AS..117..393B}. We used the automatic aperture magnitudes
determined by \sex, similar to Kron's `first moment' algorithm
\citep{1980ApJS...43..305K}. Our high galactic-latitude fields contain
a low source-density of objects (particularly in $u$'), so crowding on
the fields was not a problem and de-blending was rarely required.

\subsubsection{CFHT and DPOSS data}
\label{sec:cfht-imaging}

\noindent
A second optical survey of SA57 was conducted using the Canada France
Hawaii Telescope (CFHT) and the CFH12k instrument, a large mosaicing
camera with an imaging area of 42$\arcmin$$\times$28$\arcmin$
\citep[see][]{2000SPIE.4008.1010C}. The observations were carried out
on the 25th (3/4 night) and 26th (1/4 night) May 2000 using seven
CFHT12k pointings to cover the SA57 field.  The first night was
photometric, but poor weather on the second night curtailed the
program. Each pointing comprised two 300s dithered exposures.
\citet{1992AJ....104..340L} standard star fields were interspersed
with the science observations to achieve a photometric calibration of
the data.

The data were reduced at the TERAPIX centre using the
TERAPIX\footnote{http://terapix.iap.fr/} pipeline reduction software
designed for use with the CFH12k camera according to a procedure
described in the June 2002 TERAPIX progress
report\footnote{http://terapix.iap.fr/doc/doc.html}. An image
catalogue covering SA57 was constructed using the \sex\ package and
here again we use the automatic aperture magnitudes. The catalogue is
complete to $B=24$.

Finally, we augment the two dedicated observing programs of SA57 with
data taken from the Digitised Palomar Sky Survey (DPOSS). This
provides data for both SA57 and A1367 in $g$, $r$ and $i$ Gunn filters
\citep[see][]{2002astro.ph.10298G}. The areal coverage in both fields
is 100 per cent, though the data are shallower ($g<22.5$) and of
limited photometric accuracy when compared to the Palomar/CFHT data,
hence we use the DPOSS data where no other imaging is available.  The
optical sources in the various bands are uniquely matched within a
search radius of 1.5\arcsec.  As in our earlier papers, we use a
search radius of 10\arcsec\ to assign optical counterparts from our
combined optical data to the FOCA sources.  The problem of multiple
optical counterparts is addressed in
Section~\ref{sec:search-more-optical-counters}.

\subsection{Comparing the magnitude systems}
\label{sec:comp-magn-syst}

To consistently inter-compare these new photometric data, we place all
the magnitudes onto a system with the same calibration zeropoint. Our
imaging data comprises four different calibration systems: The FOCA
system \citep[e.g.][]{1992A&A...257...24M}, the Vega system
\citep[e.g.][]{1953ApJ...117..313J,1995PASP..107..945F}, the AB system
\citep[e.g.][]{1974ApJS...27...21O}, and the Gunn system
\citep[e.g.][]{1983ApJ...266..713O}. We compute conversions between
the different systems using the SEDs of Vega, BD+17\degr4708 (for Gunn
magnitudes), and appropriate filter and CCD response curves.  Because
of the advantages of a system where magnitudes are easy to interpret
physically, we align all of our magnitudes onto the AB system, defined
as:

\begin{equation}
  \label{eq:ab_system}
    m_{\nu}=-2.5\log (F_{\nu}) - 48.60
\end{equation}

\noindent
where $F$ is the flux in \ergscmhz. For convenience, we list the
conversion values between the various original systems to the AB
system in Table~\ref{tab:filter_conv}.

\begin{table}
  \centering
  \caption{Offsets used to convert the magnitudes to the AB system.}
  \begin{tabular}{lcccc}
Filter & \leff & \multicolumn{3}{c}{Conversion from}\\
& (\AA) & ${\rmn{foca}}\rightarrow {\rmn{AB}}$ & ${\rmn{vega}}\rightarrow {\rmn{AB}}$ & ${\rmn{gunn}}\rightarrow {\rmn{AB}}$\\
\hline
FOCA-UV      & 2015 & +2.25 &        &      \\
Palomar-$u$' & 3603 &       &  +0.92 &      \\
CFHT-$B$     & 4407 &       &  -0.10 &      \\
DPOSS $g$    & 5330 &       &        & +0.02\\
Palomar-$r$' & 6282 &       &  +0.17 &      \\
DPOSS $r$    & 6908 &       &        & -0.19\\
\hline
  \end{tabular}
  \label{tab:filter_conv}
\end{table}

\section{Optical Properties}
\label{sec:optical-properties}

\noindent
In this section we examine the photometric properties of the galaxy
sample. We first describe our procedures for assigning optical
counterparts to the FOCA source list, and then correct all of the
observed magnitudes for the effect of internal dust extinction. The
new optical coverage also allows us to estimate aperture corrections
for our spectral measures, as well as a cross-check on the accuracy of
the flux calibration of our spectra.  We then continue with an
exploration of the UV-optical properties, using the new data to derive
improved galaxy types, $k$-corrections and luminosities compared to
our earlier analyses.

\subsection{FOCA\,--\,optical counterparts}
\label{sec:search-more-optical-counters}

\noindent
One question remaining from our earlier papers concerned the number of
FOCA detections with no apparent counterpart in optical data, which
would imply either blue $UV-B$ colours or a number of false detections
in the FOCA dataset. The optical imaging now available to us allows us
to address this issue in a more comprehensive manner, as the depth of
our $B$-band data is $\sim2$ magnitudes deeper than the optical
dataset used in S2000. This magnitude limit enables us to detect all
the optical counterparts to the FOCA source list. We cross-correlate
the combined optical $u'Br'$ catalogue with the FOCA source catalogue
using a search radius of 10\arcsec\ (see e.g. T98 or S2000 for details
of the astrometric precision of the FOCA data).  For each FOCA object,
we record the number of matches, and assign each the nearest optical
detection.

The number of multiple counterpart cases (where more than one optical
detection lies within the search radius) is increased compared to
S2000; however the number of UV detections with no optical counterpart
is smaller: 50 per cent of the FOCA sources have multiple counterparts
whilst less than 6 per cent have none. This 6 per cent thus represents
a reasonable lower limit to the false detection rate in the FOCA
source list, as the magnitude limit of the new data allows us to
detect FOCA counterparts $\sim1$ magnitude bluer in $UV-B$ than the
bluest unreddened model starburst galaxy SED.  The extra number of
counterparts we find here are therefore most likely chance
associations resulting from the higher surface density of objects in
the fields: the brightest optical counterpart is also the closest in
85 per cent of the multiple counterpart cases. Where the brightest
optical detection is more than 3 magnitudes brighter then the next
brightest counterpart, we assume that the contribution to the UV flux
of the fainter object is negligible and ignore it in our analyses.  In
our following analyses, `multiple optical counterparts' galaxies are
defined as those FOCA objects with two optical counterparts within
$B$-band measures within 3 magnitudes of each other.

\subsection{Dust and $k$-corrections}
\label{sec:dust-corrections}

There are a variety of extinction laws (e.g., Milky Way, LMC, SMC)
that could be used to apply the dust corrections factors derived from
the Balmer decrements to the UV and optical magnitudes, giving quite
different extinction corrections, particularly at UV wavelengths.  The
reddening of the UV continuum depends sensitively on the geometrical
details of the dust-star-gas mix, with the reddening of the stellar
continuum likely to be different to the obscuration of the ionised gas
(derived empirically from the nebular emission via the Balmer
decrement), as the stars and gas may occupy different areas within a
galaxy with differing dust covering factors \citep[see
e.g.][]{1988ApJ...334..665F,1994ApJ...429..582C,1999A&A...349..765M}.
As in our previous work, we use the \citet{2000ApJ...533..682C}
prescription, derived empirically from studies of local starburst
galaxies, to `correct' our optical fluxes.
\citet{2000ApJ...533..682C} find that the colour excess of the stellar
continuum, \ebmvs, is related to the colour excess of the ionised gas,
\ebmvg, as $\ebmvs=0.44\,\ebmvg$.

We apply this reddening prescription to every filter passband in use
in our survey using \ebmvg\ derived from the Balmer decrement. Where
the \hb\ line is not detected, we use the SFR-dependent correction
derived for this sample of galaxies presented in
\citet{2001ApJ...558...72S} using an empirical SFR derived from the
\ha\ line (where both the \ha\ and \hb\ lines are absent, we use the
median \ebmvg =0.3 representative of the whole sample). It is becoming
apparent that luminosity-dependent dust corrections of this nature are
not valid for star-formation across all Hubble types. However, for the
mainly starburst or strongly star-forming galaxies typical in this
sample, the luminosity dependent correction still appears an excellent
approximation \citep{2003astroph0307175A}.

We convert our observed (and dust-corrected) magnitudes, $m$, in the
various passbands into rest-frame magnitudes by writing the corrected
absolute magnitude, $M$, at $z=0$ as

\begin{equation}
  \label{eq:mag_kcorr}
M=m_{\rmn{obs}}-5\log(\frac{D_L(z)}{10\,\mbox{pc}})-k(z)
\end{equation}

\noindent
where $D_L$ is the luminosity distance at redshift $z$,
$m_{\rmn{obs}}$ is the observed apparent magnitude, and $k(z)$ is the
$k$-correction \citep[see, for
example,][]{1988ApJ...326....1Y,1997A&AS..122..399P}.  We neglect any
redshift-evolution in the physical properties of the galaxies in this
calculation.

The magnitude of the $k$-corrections for template galaxy spectra using
the range of the filters in this study strongly depends on galaxy
type, with early-type galaxies (ellipticals and S0s) possessing larger
$k$-corrections in the optical passbands than galaxies with flatter
spectra. However, $k$-corrections in the FOCA passband are small with
the largest $k$-corrections arising from elliptical and S0 type
spectra, unlikely to be the predominant galaxy type in this
UV-selected sample of objects.

We assign each object in our survey a $k$-correction by fitting the
observed magnitudes to a set of model galaxy templates of varying
type. Though we could do this by correlating our observed spectra with
a set of template spectra, either via cross-correlation techniques
\citep[e.g.][]{1995AJ....110.1071C,1995AJ....110.1602Z,1997MNRAS.285..613H}
or using a principal component analysis \citep[see][and references
therein]{1999MNRAS.308..459F,2002MNRAS.333..133M}, this approach
requires spectra of a high S/N and wide wavelength coverage, which is
not always the case in the spectra of our fainter objects. Instead, we
use the popular technique of fitting a series of template SEDs to the
broadband colours available for each galaxy and use the best-fitting
SED for the estimation of $k$-corrections in each filter.  This was
also the approach used in T98 and S2000 though based on only a single
$UV-b$ colour and using POSS-APM magnitudes, which have proved
unreliable for bright galaxies. Here, we perform a more accurate
calculation with the improved photometric coverage.

We determine our $k$-corrections by fitting each colour for a given
galaxy to SED set, linearly interpolating between the SED types.
These fits are combined by weighting inversely by the variance in the
observed colours to obtain a mean best-fitting SED for each galaxy.
For the dust-corrected colours we compare to the SEDs of
\citet{1997A&AS..122..399P} with 6 galaxy classes ranging from
early-type galaxies (E/S0) to late-type spirals (Sc/Sd) supplemented
with two starburst (SB) models, generated by superimposing a starburst
on a passively evolving system (see T98 and S2000 for details).  For
our uncorrected colours we use the SEDs of
\citet{1980ApJS...43..393C}, with 4 classes of SEDs from elliptical
(E) to irregular (Im).

\begin{figure}
  \centering
  \includegraphics[width=80mm]{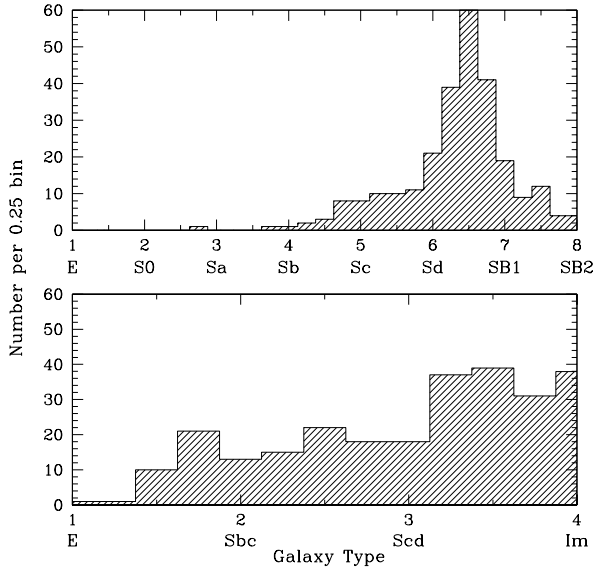}
  \caption{
    The distribution of galaxy types ($T$) in our survey. UPPER: $T=1$
    corresponds to an E SED, $T=8$ corresponds to an SB2 SED. SEDs
    from \citet{1997A&AS..122..399P}. LOWER: $T=1$ corresponds to an E
    SED, $T=4$ corresponds to an Im SED. SEDs are from
    \citet{1980ApJS...43..393C}. }
  \label{fig:galaxy_types}
\end{figure}

\begin{figure}
  \centering
  \includegraphics[width=80mm]{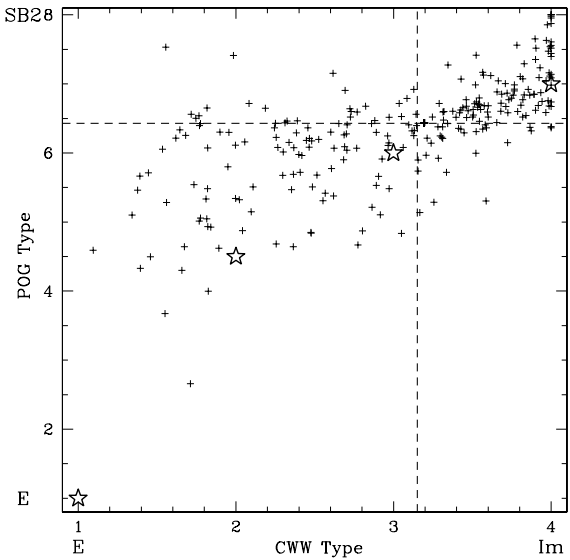}
  \caption{
    Comparison of the galaxy types derived from the two SED sets, with
    the CWW SED types on the $x$-axis and the Poggianti SED types on
    the $y$-axis, demonstrating the agreement between the galaxy types
    derived from the two sets of SEDs, as well as before and after
    dust corrections have been applied. The stars show locations where
    the two types approximately correspond.}
  \label{fig:compare_types}
\end{figure}

In Fig.~\ref{fig:galaxy_types}, we present the results of the fitting
process, showing the distribution of galaxy types ($T$) using each set
of SEDs, with $T=1$ representing the reddest SED. The different
classes are summarised in Table~\ref{tab:galaxy_classes}.
Fig.~\ref{fig:galaxy_types} shows the majority of the galaxies are
types Sb or later, with very few early-type objects.  There is a good
correspondence between the derived types (see
Fig.~\ref{fig:compare_types}), with the \citet{1997A&AS..122..399P}
derived types on average slightly bluer, possibly due to the inclusion
of dust corrections in the derivation of these types. The mean types
in each case are $T=3.15$ for the CWW SEDs ($\simeq$~Sd) and $T=6.43$
for the Poggianti SEDs ($\simeq$~Sd). No galaxies appear redder than
the model SEDs, though at the blue end, 3 of the galaxies are bluer
than the bluest CWW SED, whilst 2 are bluer than the bluest SB2
Poggianti SED.  This points to a large population of actively
star-forming or starbursting galaxies in the sample, as expected.

One caveat is that these galaxy types (and hence $k$-corrections) are
based on the \textit{weighted average} of (up to) four colours, and
hence will not be sensitive to `peculiar' colours in a single pair of
filters where the other colours of that galaxy are `normal'. This
might occur (for example) in a galaxy experiencing a sudden burst of
star-formation; the $UV-B$ colours might appear very blue whilst the
$B-r'$ or $r-i$ colours could be almost unchanged from their pre-burst
values. We return to this issue in later sections.

\begin{table}
  \centering
  \caption{The correspondence between the different galaxy classes.}
  \label{tab:galaxy_classes}
  \begin{tabular}{ccccccccc}
SED set   & E & S0& Sa& Sb& Sc& Sd& Im/SB1& SB2\\ 
\hline
Poggianti & 1 & 2 & 3 & 4 & 5 & 6 & 7     & 8\\
CWW       & 1 &   &   & 2 &\multicolumn{3}{c}{3~~~~~~~~~4}&  \\
\hline 
  \end{tabular}
\end{table}

\subsection{Aperture Corrections}
\label{sec:aperture-corrections}

In S2000, we saw that \ha\ derived SFRs were generally smaller than
UV-based SFRs by factors of around 3\,--\,4.  Potential (non-physical)
explanations of this effect could be related to uncertainties in the
flux calibration of the spectral data, or if the 2.7\arcsec\ diameter
fibres used on the WYFFOS spectrograph were sampling only a small
fraction of the total galaxy light, and hence generating an aperture
mismatch when compared to the FOCA magnitudes. In S2000, we found
little significant trend in the ratio of \ha/UV light with redshift
(and hence apparent size), as would be expected if aperture
corrections were not a significant source of uncertainty. However, the
new $B$ and $r$'-band data allow us to address any remaining
uncertainties, as well as to extend the number of objects in the
survey with a calibrated \ha\ measure.

For every FOCA $B$ and $r$' counterpart, we measure from the imaging
data the magnitude of each object inside a radius of 2.7\arcsec (the
WYFFOS fibre radius) using \sex. We convolve the flux-calibrated
spectrum with the filter+CCD response of the $B$/$r$' filters to
generate corresponding `spectral magnitudes'. We find that the
spectral $B$-$r$' colours typically agree with the optical 2.7\arcsec\ 
$B$-$r$' colours to within 10-20 per cent, indicating little
wavelength-dependent uncertainty in the flux calibration. We then
calculate aperture corrections by comparing the $r$' spectral
magnitudes with the total $r$' magnitude (as measured by \sex),
deriving correction factors by which we scale the \ha\ fluxes. We see
minor evidence for a magnitude dependent effect, with some apparently
brighter, and likely nearer with a greater apparent size, objects
requiring larger correction factors. We exclude galaxies which require
very large corrections from our sample and the future analyses in this
paper. The median factor for the survey is 1.26, i.e.  0.25\,mag. For
the small number of galaxies without any $r$' information this median
correction is applied. This median factor compares well with that
derived by \citet{1998ApJ...495..691T}, who find a mean correction of
0.52\,mag using $V$ band magnitudes for the CFRS galaxies, but with
smaller 1.75\arcsec\ slits, thus accounting for their slightly larger
aperture corrections.

The $r$'-band data also allow us to apply a `flux calibration' to \ha\ 
fluxes which were taken during an observing run for which no standard
stars were observed (detailed in T98), following a similar procedure
as with the aperture corrections. This adds a further 17 objects to
our \ha\ sample.

These corrections assume that the \ha\ emission in the galaxy is a
uniform distribution, and that the galaxy is not dominated by either
nuclear or disc star-formation. Whilst these uncertainties may be
significant for local nearby objects (with a larger apparent size), by
redshifts of 0.05 (i.e. for most of the galaxies in our sample) these
effects are likely to be very small.

\subsection{UV/Optical colours}
\label{sec:uvoptical-colours}

One of the most puzzling features of the FOCA galaxies presented in
our earlier papers was the presence of a sub-sample of galaxies with
extremely blue $UV-B$ colours, bluer than most typical starburst SEDs.
The improved optical data that we have assembled now allows us to
examine the SEDs of these objects across a wider wavelength range.

\begin{figure}
  \centering
  \includegraphics[width=80mm]{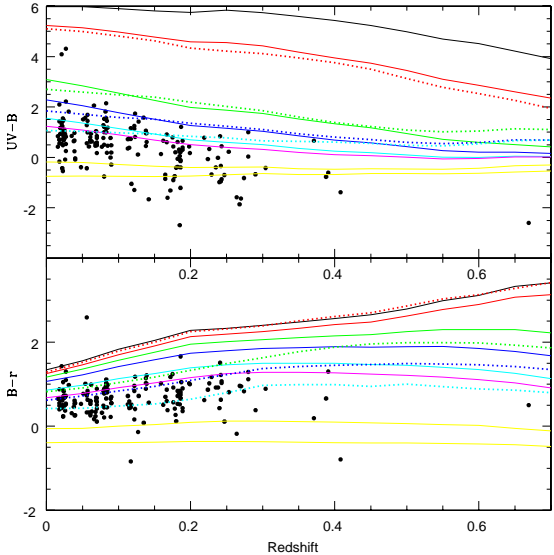}
  \caption{
    Colour-redshift distributions for the unique optical counterpart
    cases only. UPPER: $UV-B$, LOWER: $B-r$.  The solid lines show the
    model predictions for the \citet{1997A&AS..122..399P} set of SEDs,
    the dotted lines are the colours derived from the CWW SED's (see
    text for details).  The galaxies appear slightly bluer in $UV-B$
    compared to the model SEDs than for $B-r$; additionally in $UV-B$
    there is a small population of objects with colours bluer than the
    bluest model SED. The colours are in the AB system and are
    corrected for dust extinction.}
  \label{fig:uv_b_r}
\end{figure}

With the new optical magnitudes, 8 and 5 per cent of the UV galaxies
with a unique optical counterpart (slightly more when including
multiple counterpart cases) are found to be bluer than the bluest
Poggianti SED in $UV-B$ and $UV-g$ respectively (where the $g$ data
also includes the A1367 field). Before dust correction, these
fractions are $\simeq4$ and $\simeq2$ per cent respectively. At longer
wavelengths, the galaxies have more `normal' colours typical of the
mean SED of the sample (see Fig.~\ref{fig:uv_b_r}).  As mentioned
earlier, this might occur in galaxies experiencing a sudden burst of
star-formation (see Section~\ref{sec:modelling}). A second possibility
is that that hot Wolf-Rayet (WR) stars could be responsible for the UV
excess as suggested by \citet{2000ApJ...540L..83B}. We examined this
suggestion in detail in \citep{2002MNRAS.330...75C}, studying the
co-added spectrum of the bluest objects in the sample, but could
identify no WR spectral features in the spectrum of galaxies with
extreme $UV-B$ colours.

\section{Star-Formation properties of the Sample}
\label{sec:star-form-prop}

We now turn to the main focus of this paper, continuing our
investigations into the star-formation properties of the survey begun
in S2000, where we examined the relationship between \ha\ and UV
derived SFRs. We found that this relationship possessed a scatter and
non-unity best-fitting slope which could not be satisfactorily
explained in scenarios assuming constant or other simple SFHs. We now
use our new data to investigate these effects with increased
precision.

\subsection{Uncertainties in converting luminosities into star-formation rates}
\label{sec:conv-lumin-into}

Our first step is to construct a framework within which we can
transform our measured luminosities for each of the star-formation
diagnostics into SFRs.  As has been covered extensively in other
studies, there are a remarkable number of free parameters in these
models -- the initial mass function (IMF), the stellar metallicity,
and the choice of SED libraries, amongst many others \citep[see][for a
discussion of the issues]{1998ARA&A..36..189K}.  We derive conversion
factors in a self-consistent manner using the \pegase2\ spectral
synthesis code \citep{1997A&A...326..950F,1999astro.ph.9912179}. The
three star-formation diagnostics available in this study are

\begin{itemize}
  
\item \textit{\ha\ luminosity (\lha)}, originating from re-processed
  ionizing radiation at wavelengths $\lambda<912$\,\AA\ produced by
  the most massive ($>10\,\rm{M}_{\odot}$), short-lived
  ($\simeq20\,\rm{Myr}$), OB-type stars.  Accordingly, \ha\ emission
  is a virtually instantaneous star-formation measure,reaching a
  constant level after $\simeq10\,\rm{Myr}$ in a constant SFH.
  However, it is very sensitive to the form of the IMF due to the
  strong dependence on massive stars.  Most calibrations assume case-B
  recombination \citep[a comprehensive treatment of its use can be
  found in][]{2001MNRAS.323..887C}.
  
\item \textit{UV 2000\,\AA\ continuum luminosity (\luv)}, originating
  from stars spanning a range of ages (and hence initial masses),
  including some post-main sequence contribution. Hence, any \luv\ to
  SFR calibration is dependent on the past history of star-formation,
  introducing a significant uncertainty when interpreting UV
  observations of star-forming galaxies. Dust corrections are also
  important in the UV, and a reliable correction can be difficult to
  achieve \citep[e.g.][]{2002ApJ...577..150B}.
  
\item \textit{$u$' 3600\,\AA\ continuum luminosity (\lu)}, with a
  similar physical origin as \luv. The principle disadvantage is that
  $u$' luminosities are contaminated by older, less massive stars to a
  greater extent than the UV-2000\,\AA\ data (given the longer
  wavelength), and this makes the calibration less certain than with
  shorter wavelength data. The advantage of using longer wavelengths
  is that dust corrections are smaller.
\end{itemize}

Using \pegase2, we investigate the dependence of the three diagnostics
to the principle free parameters in the spectral synthesis codes,
varying each to determine the spread in the conversion values. We
investigate metallicities of $Z=0.0004$, 0.004, 0.008, 0.02
($=Z_{\odot}$) and 0.05, together with the IMFs of
\citet{1955ApJ...121..161S}, \citet{1998simf.conf..201S} and
\citet{2001MNRAS.322..231K} using mass ranges of
$0.1\rightarrow100\,\msun$, $0.1\rightarrow120\,\msun$,
$0.1\rightarrow50\,\msun$, $1.0\rightarrow100\,\msun$ and
$5.0\rightarrow100\,\msun$.

\begin{table}
  \centering
  \caption{Default conversions between luminosity and SFR}
  \begin{tabular}{lll}
Tracer & \multicolumn{2}{c}{Luminosity for $\rmn{SFR}=1\,\mperyr$}\\
& \ergsa & \ergshz\\
\hline
\ha & $1.22\times10^{41}$ \ergs & --- \\
UV 2000\,\AA & $5.75\times10^{39}$ & $7.78\times10^{27}$\\
$u$' 3600\,\AA & $1.71\times10^{39}$ & $7.40\times10^{27}$\\
\hline
  \end{tabular}
  \label{tab:sfr_convert}
\end{table}

\begin{figure}
  \centering
  \includegraphics[width=80mm]{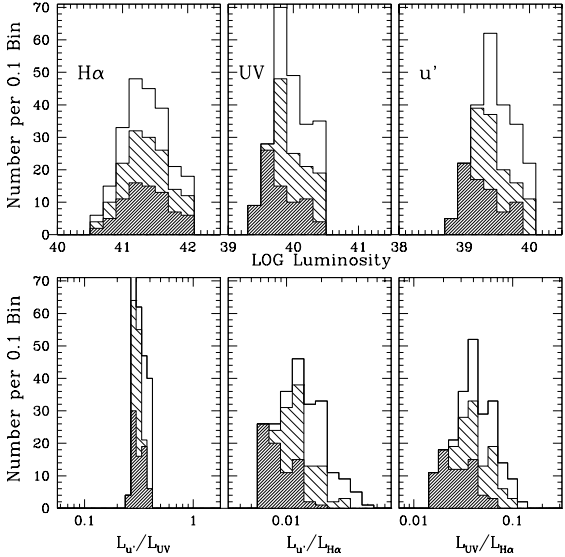}
  \caption{The distribution of the conversion values for the \ha, UV and $u$'
    diagnostics (UPPER PANEL), and the distribution of the ratio of
    those values (LOWER PANEL), generated across the full range of
    metallicities, IMFs and mass ranges (see text for full details).
    The luminosities are those expected from star-formation rates of
    $1\,\mperyr$. The shaded histograms refer to conversions taken at
    10 (heavy shading), 100 (light shading) and 1000\,Myr (no shading)
    in a constant SFH.  The figure demonstrates how the ratios
    involving \ha\ and UV (or $u$') evolve according to the time after
    the onset of star-formation, whereas the ratio of $u$'/UV has
    little time dependence.}
  \label{fig:sfrs_graph}
\end{figure}

The uncertainties in the star-formation estimates due to these
parameters are shown graphically in Fig.~\ref{fig:sfrs_graph} as the
distribution of \lha, \luv\ and \lu\ and the distribution of the
ratios of \luv/\lha, \lu/\lha\ and \lu/\luv\ for a constant SFR of
1\,\mperyr.  There is a large spread in the conversion values for all
diagnostics when considering the full range of IMFs and metallicities
(around 1.5 orders of magnitude), with the importance of the timescale
of recent star-formation for UV and $u$' diagnostics clear.  However,
the scatter when considering the ratios of the diagnostics is much
smaller, again with a strong time dependence. Though varying
metallicities or IMFs can have a large effect on individual conversion
values, the effects on the ratios is considerably smaller. Our default
conversion values (listed in Table~\ref{tab:sfr_convert}) assume solar
metallicity, a Salpeter IMF with a mass range of 0.1 to 100\,\msun,
and are taken 100\,Myr into a constant SFH.

\subsection{Comparing star-formation diagnostics}
\label{sec:comp-star-form}

\begin{figure*}
  \centering
  \includegraphics[width=80mm]{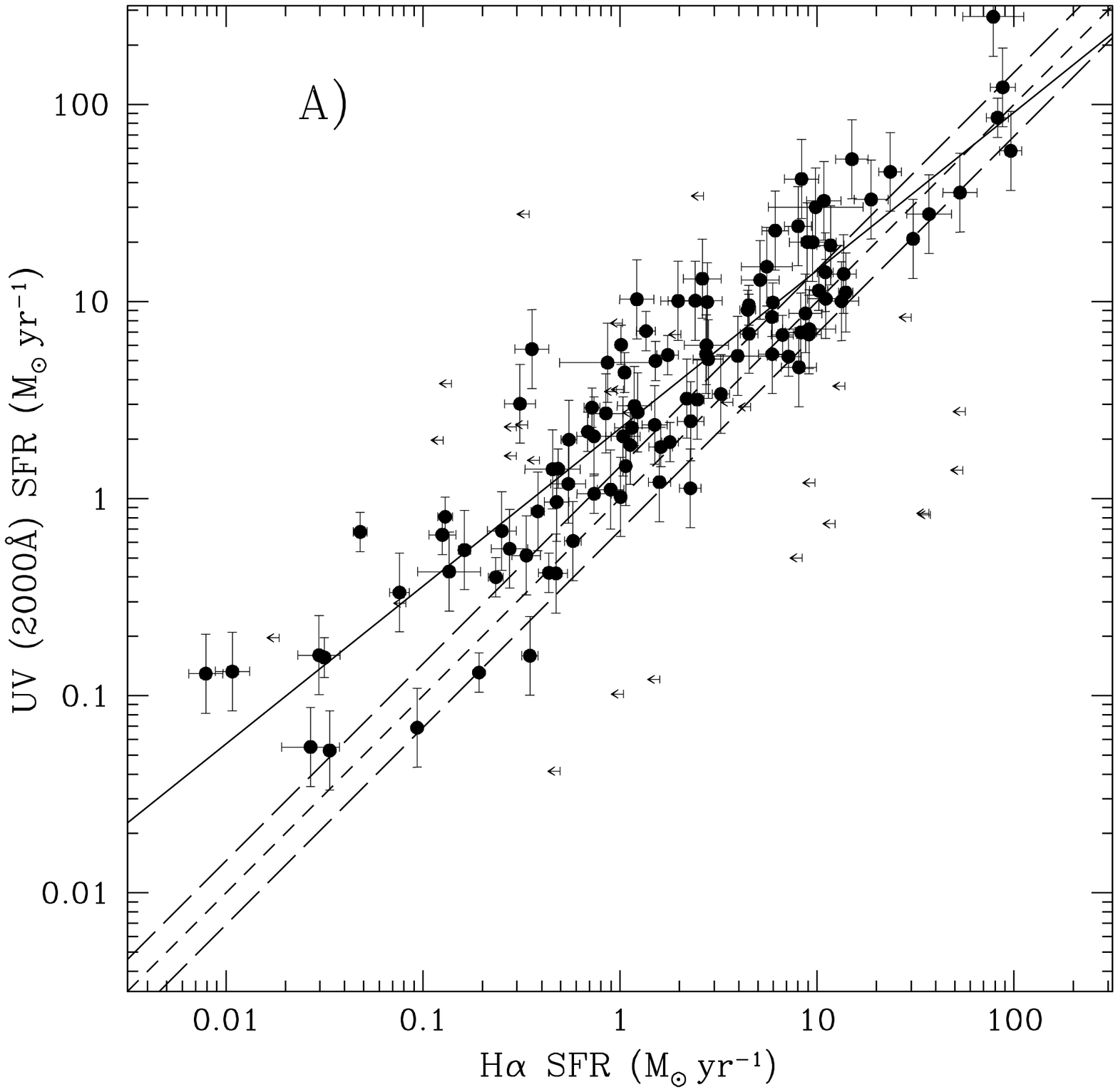}
  \includegraphics[width=80mm]{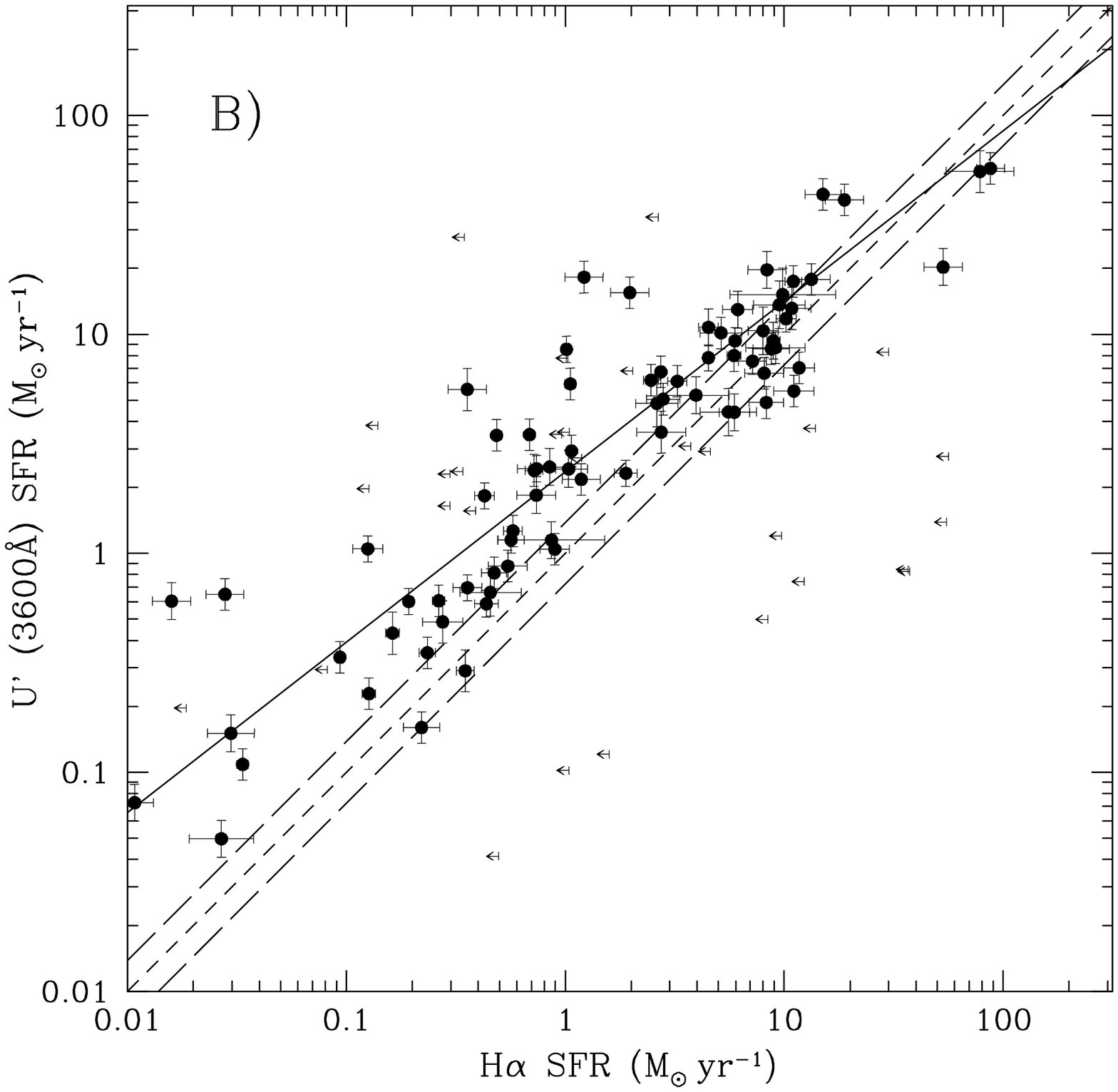}
  \includegraphics[width=80mm]{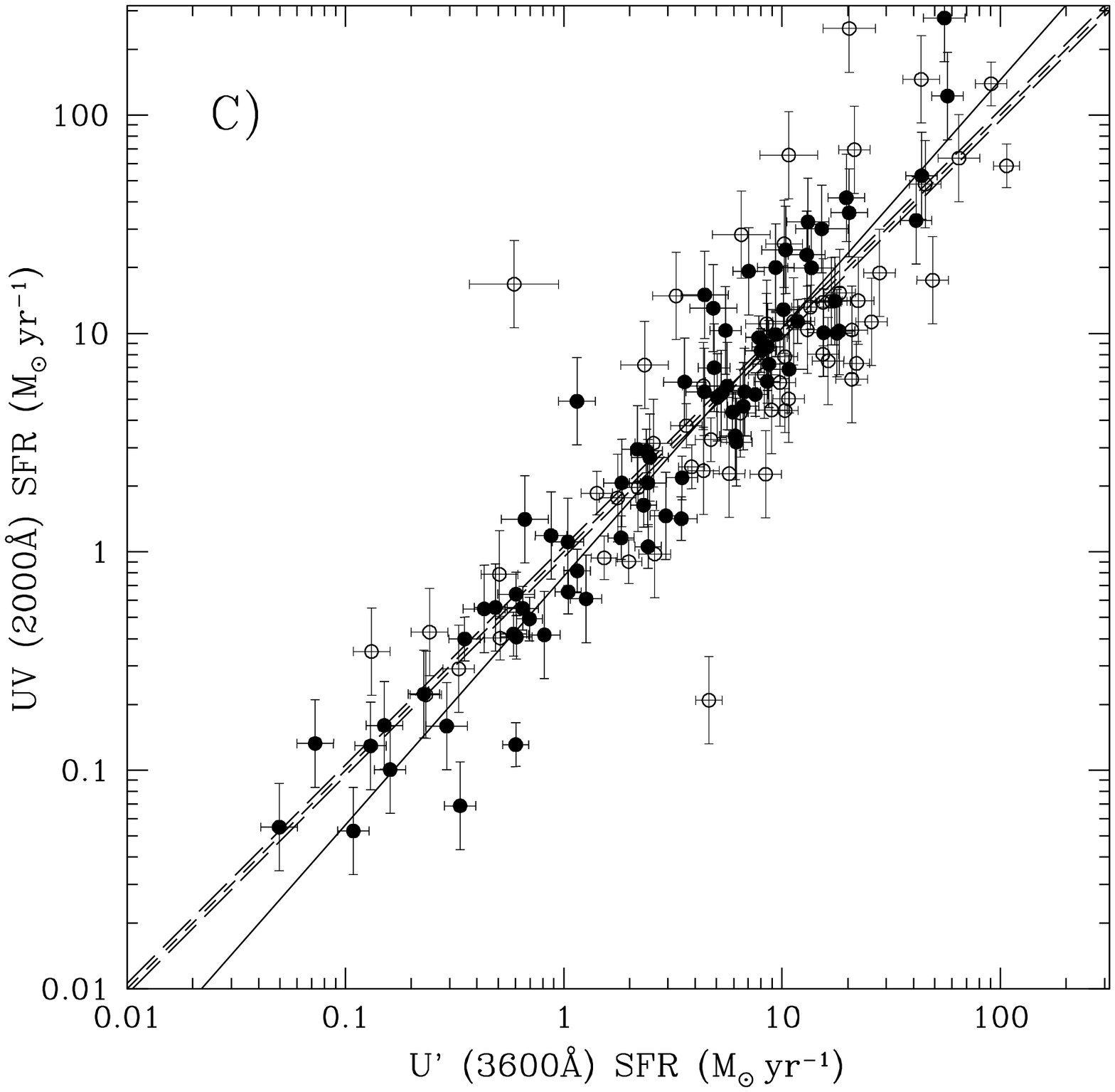}
  \includegraphics[width=80mm]{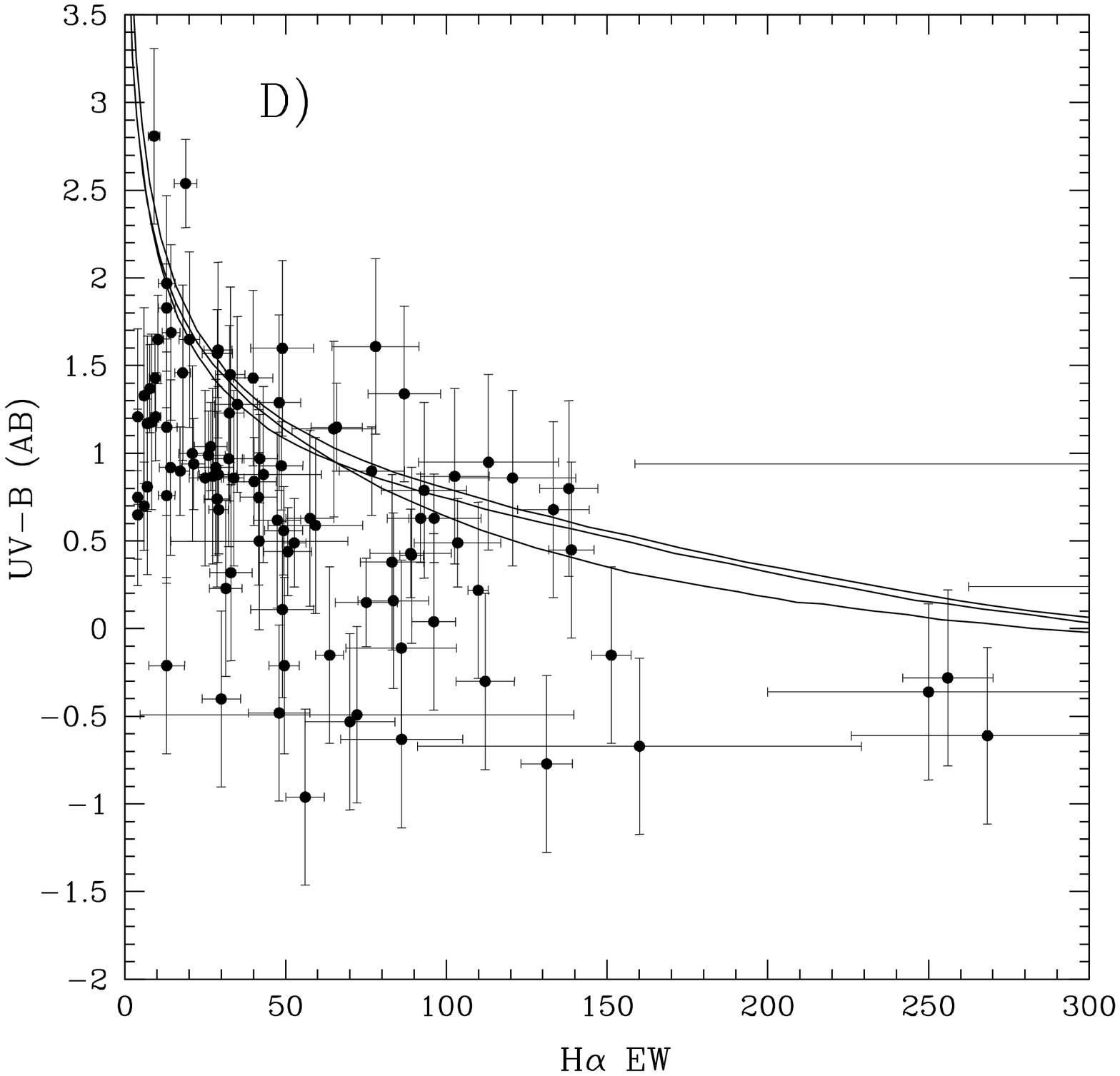}
  \caption{
    A comparison of A) UV\,2000\,\AA\ and \ha\ SFRs, B)
    $u$'\,3600\,\AA\ and \ha\ SFRs, C) UV\,2000\,\AA\ and
    $u$'\,3600\,\AA, and D) the UV-B colour and the \ha\ EW. All
    measures are dust corrected (and for \ha\ aperture corrected) as
    described in the text. In the UV\,2000\,\AA\,--\,$u$'\,3600\,\AA
    plot, galaxies possessing \ha\ emission are marked as filled
    circles, those without as empty circles. The SFRs are calculated
    at 100\,Myr into a constant SFH. The short-dashed lines denote an
    equality of SFRs and the long-dashed lines show typical
    uncertainties in the position of the short-dashed lines due to
    IMF/metallicity effects.  Solid lines indicate weighted
    least-squares best-fits. In panel D), the overlaid lines
    correspond to the relation expected in a exponentially declining
    SFH for the three IMFs discussed in the text.}
  \label{fig:compare_sfrs}
\end{figure*}

We present the relations between the different diagnostics in
Fig.~\ref{fig:compare_sfrs}, showing comparisons between
dust-corrected \ha, UV and $u$' SFRs, overlaid with lines expected
assuming constant star-formation and associated uncertainties in the
position of this line derived from the analysis of the last section.
Also shown are the error-weighted least-squares best-fits to the data.
The \ha-UV plot is an updated version of fig.~13 in S2000 with the new
aperture and $k$-corrections applied, together with some new measures
and upper limits for \ha. Linear correlation (Pearson's $r$) and
Spearman rank-order correlation statistics, together with the weighted
least-square best-fit equations and associated $\chi^2$, are presented
in Table~\ref{tab:allsfr_stats}. We do not plot the relations before
dust correction, and instead list the same statistical data for these
samples in Table~\ref{tab:allsfr_stats} as a comparison and so the
effect of our dust-correction prescription can be seen.

 \begin{table*}
  \centering
  \caption{
The results for the various statistical tests for Fig.~\ref{fig:compare_sfrs}.}
  \label{tab:allsfr_stats}
  \begin{tabular}{cccccl}
\hline 
$\log(\rmn{SFR})$ & $N$ & Pearson & Spearman & $\chi^2$ & Equation\\
Relation  & & & & &\\
\hline
\multicolumn{6}{l}{\textbf{Corrected for dust:}}\\
\ha\,--\,UV & 106 & 0.918 & 0.916 & 348 & $\log(\rmn{SFR}_{\rmn{uv}})=0.79\times\log(\rmn{SFR}_{\ha})+0.36$\\
\ha\,--\,$u$' & 78 & 0.909 & 0.906 & 793 & $\log(\rmn{SFR}_{u'})=0.78\times\log(\rmn{SFR}_{\ha})+0.37$\\
$u$'\,--\,UV  & 78 & 0.952 & 0.944 & 134 & $\log(\rmn{SFR}_{\rmn{uv}})=1.13\times\log(\rmn{SFR}_{u'})-0.11$\\
\multicolumn{6}{l}{\textbf{Uncorrected:}}\\
\ha\,--\,UV & 106 & 0.889 &  0.887 & 391 & $\log(\rmn{SFR}_{\rmn{uv}})=0.72\times\log(\rmn{SFR}_{\ha})+0.15$\\
\ha\,--\,$u$' & 78 & 0.901 & 0.903 & 892 & $\log(\rmn{SFR}_{u'})=0.75\times\log(\rmn{SFR}_{\ha})+0.31$\\
$u$'\,--\,UV  & 78 & 0.953 & 0.945 & 143 & $\log(\rmn{SFR}_{\rmn{uv}})=1.03\times\log(\rmn{SFR}_{u'})-0.25$\\
\hline 
  \end{tabular}
\end{table*}

These plots confirm several of the trends seen in S2000. Firstly, as
in our previous studies we find correlations across three decades in
luminosity between the different diagnostics. In all cases, the
correlation coefficients are greater than 0.90 (before and after dust
correction).  Secondly, the plots involving \ha-derived SFRs show a
marked increase in scatter when compared to the $u$'--UV plot (see the
$\chi^2$ figures for the best-fits in Table~\ref{tab:allsfr_stats}).
The lines derived from the analysis of
Section~\ref{sec:conv-lumin-into} indicate that this scatter is
unlikely to be \textit{primarily} generated by varying IMFs and
metallicities.  Thirdly, the relation between H$\alpha$ and UV-derived
SFRs is complex, appearing luminosity (or SFR) dependent with the
best-fit lines possessing non-unity slopes. At higher luminosities,
the diagnostics agree well, whereas at lower luminosities the \ha\ 
diagnostic typically under-estimates the SFR when compared to both
$u$' and UV continuum measure as shown by the slopes of the
best-fitting equations in Table~\ref{tab:allsfr_stats}.  The
comparison between the UV and $u$' diagnostics shows little luminosity
effect with no offset from the default metallicity and IMF.  Finally,
Table~\ref{tab:allsfr_stats} suggests that the scatter that we see is
not a result of an inappropriate dust correction being applied; the
scatter in the relations decreases after our standard dust correction
is performed.

Further discrepancies from relations expected as a result of regular
SFHs are also seen in panel D of Fig.~\ref{fig:compare_sfrs}, which
compares $UV-B$ colours with the \ha\ equivalent widths. For a given
\ha\ EW, many of the galaxies appear bluer than would be expected in
simple SFHs. Again, simply varying the IMF does not appear able to
reproduce the effect.

As the UV and $u$'-band luminosities are measured at neighbouring
wavelengths, they have a very similar dependence on the recent SFH (or
the timescale over which the SFR varies) -- see
Fig.~\ref{fig:sfrs_graph}. The dominant cause of scatter in the
UV\,--\,$u$' relation will be due to a combination of incorrect
dust-extinction corrections together with observational uncertainties
(and the possibility of contamination by non star-forming galaxies),
rather than varying star-formation timescales.  The $u$'--UV relation
for galaxies with detected \ha\ emission (i.e.  known to be
star-forming) is tighter than for galaxies selected without regard to
\ha\ emission -- indeed, many of the galaxies which show a large
discrepancy between the UV and $u$' emission show \textit{no
  detectable \ha\ emission}. Possibly these galaxies are not actively
star-forming, with a substantial fraction of the UV and $u$' light
arising from the presence of older and evolved, possibly post
main-sequence, stellar populations.

Though the luminosity (or SFR) dependent effect that we see in the
plots involving \ha-emission could be due to a luminosity-dependent
dust extinction relation
\citep[e.g.][]{2001AJ....122..288H,2001ApJ...558...72S}, the
correction we applied in Section~\ref{sec:dust-corrections} already
does a good job of removing such effects in the radio--UV relations
for these galaxies \citep{2001ApJ...558...72S}, making it unlikely the
strength of the luminosity-dependent dust relation is under-estimated.

\subsection{The nature of star-formation in the sample}

The possibility of complex SF histories as been discussed in the past
as an explanation for the broad properties of late-type star-forming
galaxies \citep[e.g.][]{1973ApJ...179..427S}. These studies have shown
that periods of enhanced star-formation relative to a galaxy's history
as a whole are able to reproduce well the blue colours of local dwarf
galaxies, as well as the scatter that is observed in the colours of
these galaxies. In S2000, we introduced the possibility of complex SF
histories as a possible explanation for the discrepancies between \ha\ 
and UV luminosities. The results of the last section based on
additional $u$-band information, an extended optical wavelength
coverage, improved $k$-corrections, and aperture corrections on the
\ha\ fluxes, confirm these findings. In S2000, we introduced the
concept of varying SFHs as a possible explanation for our dataset, and
we now investigate this further.

The key is the time-scale upon which different diagnostics of
star-formation trace changes in a galaxy's SFH. The \ha\ luminosity
depends only on the most massive and short-lived stars, and the point
at which new stars are born at the same rate as older stars die occurs
after only a few million years.  By contrast, as UV/$u$' continuum
measures have a significant contribution from older and longer-lived
stars, it takes 100 to 1000\,Myr to reach the stage at which the
birth-rate of the stars which generate UV/$u$' emission is the same as
their death-rate. A burst or increased period of star-formation
superimposed on an otherwise regular SFH therefore affects the
different diagnostic plots in different ways.

\begin{figure*}
  \centering
  \includegraphics[width=80mm]{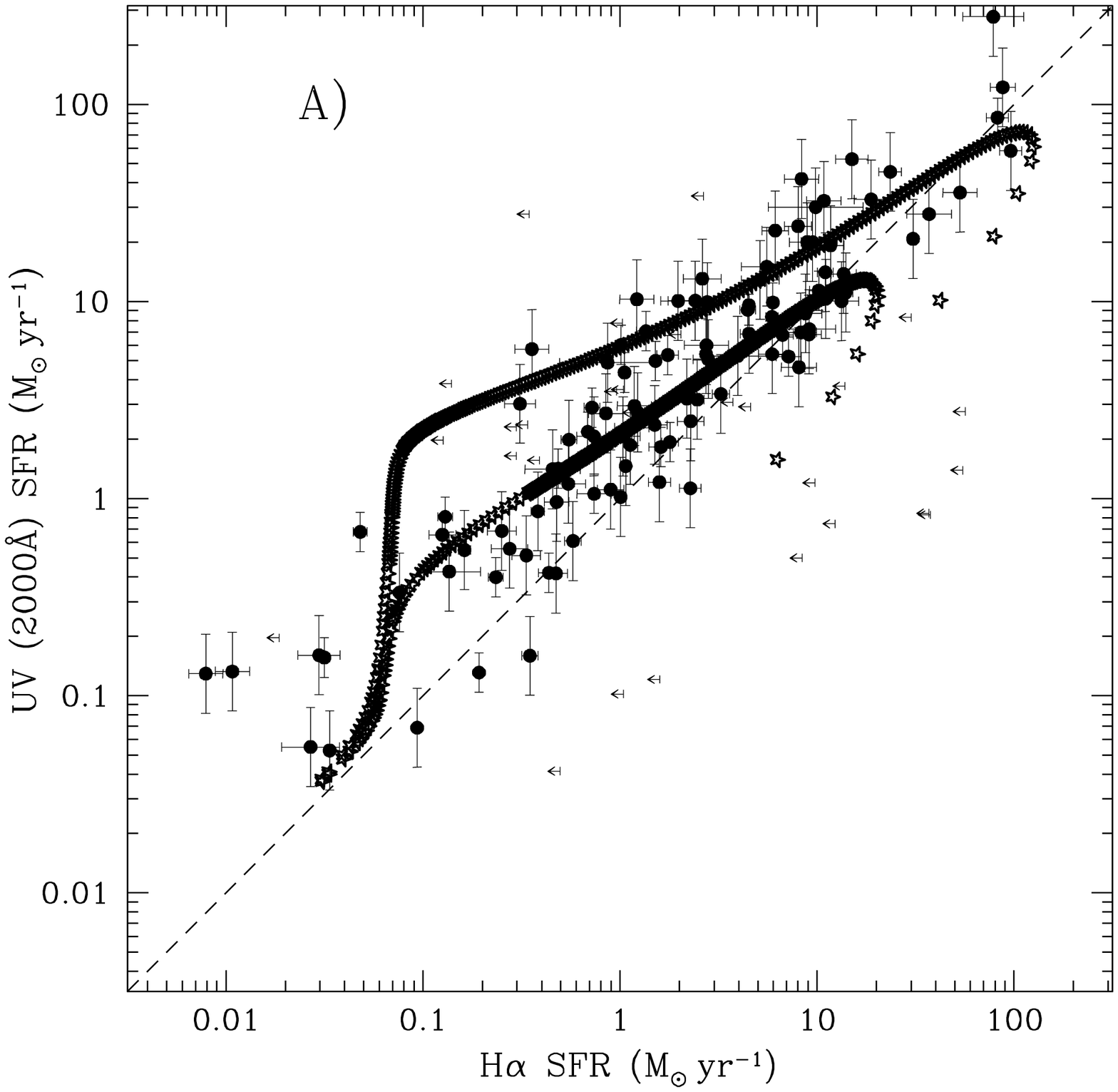}
  \includegraphics[width=80mm]{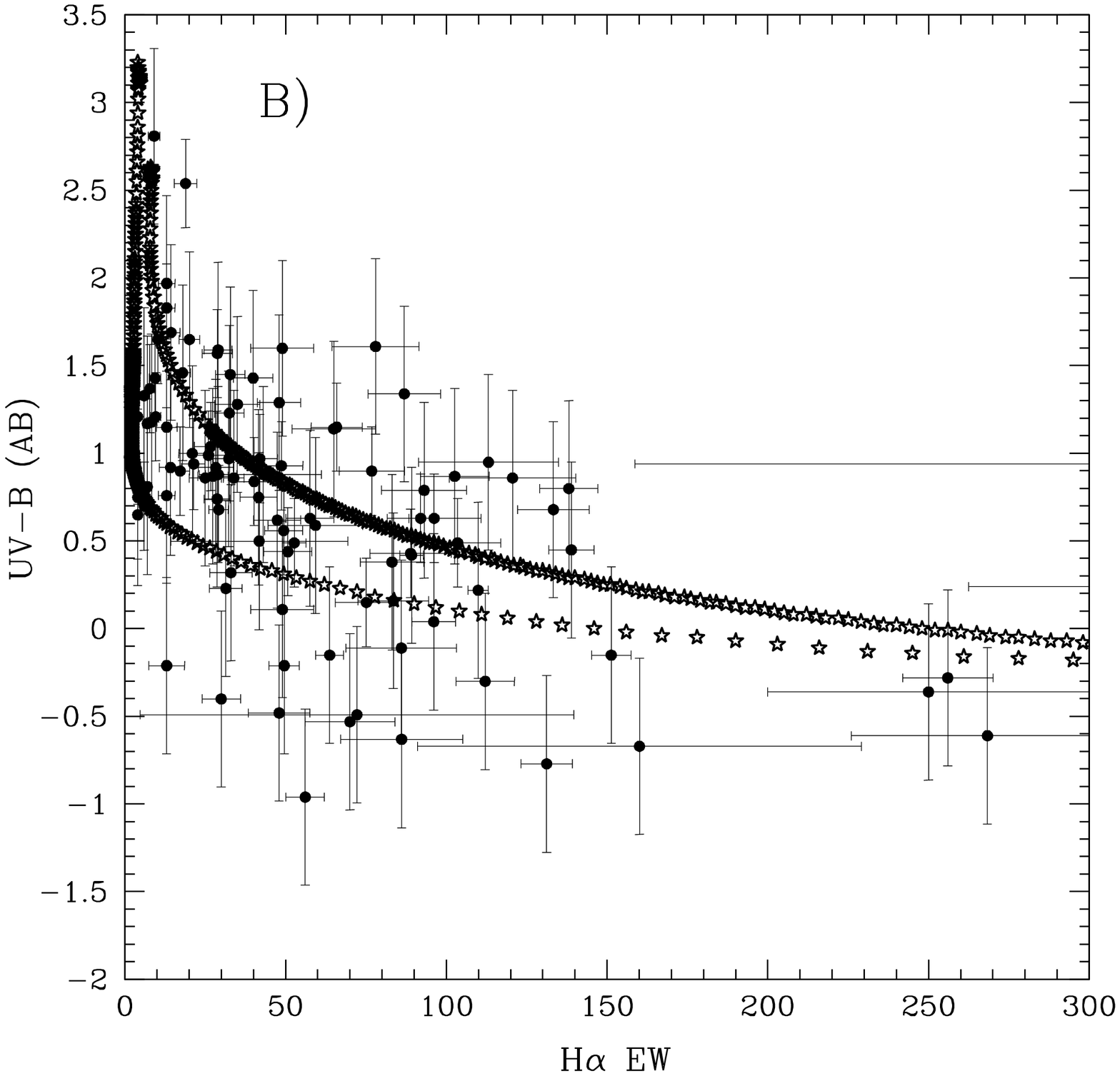}
    \caption{
      The effect of an enhanced star-formation event on A) the
      \lha\,--\,\luv\ relation, and B) $UV-B$ versus \ha\ EW. Two
      example bursts are shown: one involves 10 per cent of the galaxy
      mass in an exponentially decaying burst of e-folding time
      45\,Myr, the second 30 per cent mass with a time of 20\,Myr,
      generated using the \pegase2\ code, a standard IMF, and solar
      metallicity. The enhanced star-formation events are able to
      reproduce the scatter and offsets from constant star-formation
      scenarios seen in the data.}
  \label{fig:schematic_burst}
\end{figure*}

For the \ha\ versus UV (or $u$') plot, a star-formation event will
move a galaxy rapidly up and right on the diagnostic plot with a short
period in which the observed \ha\ SFR is higher than the observed UV
SFR as the UV light catches up with the \ha\ light. As the burst
subsequently dies away, the \ha\ rapidly decreases, whilst the UV
luminosity is retained due to the contribution from older stars.  The
galaxy describes a loop in \ha--UV (or \ha--$u$') space (see panel A
in Fig.~\ref{fig:schematic_burst}). The size of the loop depends on
the parameters of the burst: stronger (more massive) and/or shorter
bursts produce larger loops. By contrast, the effect on the UV--$u$'
relation is small as these diagnostics have a similar dependence on
the SFH. For the $UV-B$ versus \ha\ EW relation, the burst rapidly
increases the \ha\ EW and generates a bluer $UV-B$ colour. As the
burst dies away, the \ha\ returns to pre-burst levels, whilst the
$UV-B$ colour remains blue (see panel B in
Fig.~\ref{fig:schematic_burst}).

This explanation is supported by our observations and also predicts
trends we should see in our spectral diagnostics indicative of the age
of the stellar population. In galaxies which are either at the peak of
a starburst or which are undergoing a regular star-forming process
(those galaxies in which SFRs derived from \lha\ and \luv\ agree
well), we would expect small values of the Balmer decrement and high
values for the \ha\ EW due to the presence of a young stellar
population. For galaxies which are in the later stages of a burst of
star-formation, where we have excess in UV/$u$' SFRs compared to \ha,
we would expect larger Balmer breaks and weaker \ha\ EWs as the burst
population ages.

\begin{figure}
  \centering \includegraphics[width=80mm]{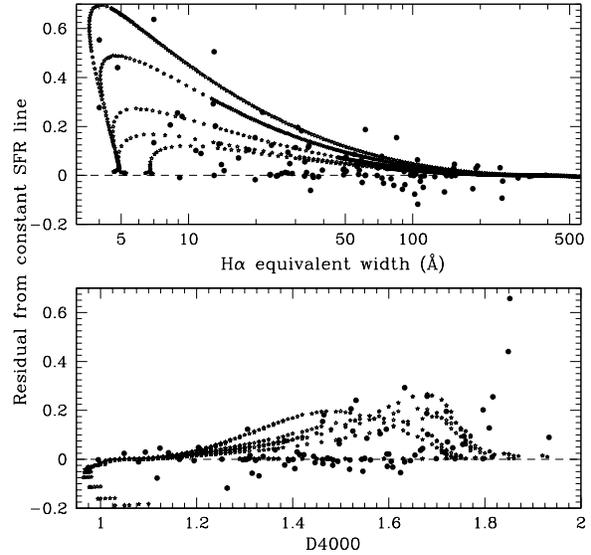}
    \caption{
      The distribution of the residuals from the constant SFR line on
      the \ha$_{\rmn{SFR}}$--UV$_{\rmn{SFR}}$ plane, as a function of
      the \ha\ EW (upper) and D4000 (lower) for the FOCA galaxies,
      shown as solid circles. The horizontal dashed lines show the
      position of constant star-forming galaxies. Overlaid stars show
      the evolution of these parameters during example starburst
      events as in Fig.~\ref{fig:schematic_burst}. Again, the
      starburst events are able to reproduce the scatter and offsets
      seen.}
  \label{fig:haew_bb_residual}
\end{figure}

We examine these parameters in Fig.~\ref{fig:haew_bb_residual},
showing the distribution of the Balmer break (D4000) and \ha\ EW
values as a function of the position of a galaxy on the
\ha$_{\rmn{SFR}}$--UV$_{\rmn{SFR}}$ diagram. As expected, we see some
evidence that those systems with the largest residuals from constant
SFR scenarios demonstrate features indicative of older stellar
populations (larger values of D4000 and smaller \ha\ EWs).

Irregular SFHs clearly help to explain the discrepancies that we see
in the relations exploring the different star-formation diagnostics,
in particular the `excess' SFRs derived from UV and $u$' measures when
compared to \ha, and the luminosity dependence of this result. We
investigate this hypothesis more quantitatively in
Section~\ref{sec:modelling}.

\section{Analysis}
\label{sec:modelling}

\noindent
We have reviewed the \textit{qualitative} evidence that many of the
low to intermediate luminosity galaxies in this data set do not
possess simple constant or smoothly declining SFHs.  This is an
important finding in the context of the interpretation of SFR measures
from large galaxy redshift surveys, our primary motivation in this
section is to \textit{quantitatively} confirm our findings. With
sample photometry that ranges from 2000\,\AA\ to $\sim7000$\,\AA\ 
(plus spectral coverage at optical wavelengths), we are able to
investigate any SFHs that are more consistent with the combined
dataset, via a simple $\chi^2$ modelling technique. We do this by
examining the form of the recent SFH in our galaxies, and then discuss
the implications for utilising the different star-formation
diagnostics for the galaxy population as a whole.

\subsection{Evidence for complex star-formation histories?}
\label{sex:fitting-galaxy-seds}

For each galaxy we have available the following (dust-corrected)
measures: i) The \ha\ luminosity, ii) The UV continuum luminosities at
2000 and 3600\,\AA, iii) The strength of the Balmer Break, D4000, iv)
The \ha\ EW, v) The $UV-B$ and $B-r$' colours.  81 of our galaxies fit
these criteria. The amount of extinction in the models is not left as
a free parameter, but is constrained observationally using the
\ha/\hb\ Balmer decrement as described earlier. Our goal is to examine
the form of the most recent star-formation event for which this is an
excellent dataset (however it is not able to provide constraints on
the SFH in the distant past). In what follows, we assume a Salpeter
IMF with mass limits at 0.1\,\msun\ and 100\,\msun, solar metallicity
and again utilise the \pegase2\ population synthesis code.

We begin by considering two simple SFHs, one representing a constant
SFH and the other an exponentially declining SFH according to

\begin{eqnarray}
  \label{eq:constant_sfh}
  \rmn{SFR}(t)&=&M_0\ \ \ (\mperyr)\\
  \label{eq:exponential_sfh}
  \rmn{SFR}(t)&=&\frac{M_0}{\tau}\exp(-\frac{t}{\tau})\ \ \ (\mperyr)
\end{eqnarray}

\noindent
In equation~(\ref{eq:constant_sfh}), $M_0$ is the SFR and the stellar
mass formed at time $t$ is $M_0\times t$; in
equation~(\ref{eq:exponential_sfh}), $\tau$ is the time constant of
the SFH (in Myr) and $M_0$ is the total mass formed at time $t=\infty$
(i.e $\int M_0/\tau \exp(-t/\tau)dt = M_0$). Small values of $\tau$
(e.g.  $\tau<5\,\rmn{Myr}$) approximate instantaneous bursts, large
values approximate continuous SFHs.

We construct a grid of galaxy spectra corresponding to the different
SFHs as a function of time, with a range of $\tau$ from 500 to
5000\,Myr for the exponential model and models ages of up to 12\,Gyr.
For each galaxy, we shift the model spectra to the observed redshift,
and calculate predictions for each of the observed parameters listed
above including nebular emission according to the prescription in
\pegase2.  We normalise each model (i.e. adjust $M_0$) so that the
predicted and observed \ha\ emission are identical.  We then calculate
a $\chi^2$ statistic for each model galaxy by comparing the observed
and predicted parameters, and find the combination of parameters which
minimise this $\chi^2$.  The probability of each $\chi^2$ is then
calculated using the incomplete gamma function for the appropriate
number of degrees of freedom in each fit.  

The key result is that the `best-fits' for 46/81 (57 per cent) of the
galaxies in the constant SFH scenarios and 37/81 (46 per cent) of
galaxies in the exponential SFH scenarios are rejected at greater than
the nominal 99.9 per cent confidence level -- i.e for just under half
the sample, the model SEDs provide poor representations of the true
physical scenario over the full range of $t$ and $\tau$. The principle
indicators of the poor fits are the $u$'/UV luminosities and the
\ha\,EWs, sensitive to the ratio of ongoing to past star-formation.

Consequently, for the 46 galaxies which are poorly fit by simple SFHs,
we extend these models by relaxing the assumption of `smooth' SFHs and
include the possibility of a burst of star-formation superimposed on
top of the underlying galaxy SFH, represented by

\begin{eqnarray}
  \label{eq:bursting_sfh}
  \rmn{SFR}(t)&=&\frac{(1-f_b)M_0}{\tau_g}\exp(-\frac{t}{\tau_g})\ \ \ \rmn{if}\;t<t_b \nonumber\\
  \rmn{SFR}(t)&=&\frac{(1-f_b)M_0}{\tau_g}\exp(-\frac{t}{\tau_g})\nonumber\\
&&\mbox{}+\frac{f_bM_0}{\tau_b}\exp(-\frac{t_b-t}{\tau_b})\ \ \ \rmn{if}\;t>=t_b
\end{eqnarray}

\noindent
with the starburst commencing at time $t_b$ and the total mass formed
$M_0$, a fraction $f_b$ formed during the starburst.  We ignore times
$t<t_b$ (simulated above), and assume $t_b=8000\,\rmn{Myr}$ and
$\tau_g=3.0\,\rmn{Gyr}$. The results are not sensitive to the choice
of these two parameters which define the underlying SED.

We fit for the combination of parameters which minimise the $\chi^2$
as before, with stellar masses estimated by integrating the SFH to the
best-fitting $t$.  The results of the fits using this framework is
quite different -- all but 6/46 of the galaxies which were poorly fit
by regular SFHs now have acceptable fits. Of these, only 3 have fits
which are excluded with a very high level of confidence with the 3
others `borderline' fits. The median best-fitting parameters are given
in Table~\ref{tab:median_bestfit_bursts}. Though these parameters do
vary widely when considering any one object, they suggest that the
most common starbursts are relatively short duration events, involving
around 8\,--\,10 per cent of the galaxy mass, and that we view these
galaxies not at the peak of a particular starburst (where \ha\ and UV
derived SFRs will be approximately equal), but instead some way into a
burst's lifetime; it is at these times that the \ha\ and UV or $u$'
luminosities are the most discrepant.

\begin{table}
  \centering
  \caption[Best-fitting parameters for `bursting' SFHs]{The median best-fitting parameters for the form of the bursts of star-formation}
  \begin{tabular}{cc}
Parameter & Median value\\
\hline
$\tau_b$ & 23.1 \\
$t_b$    & 125  \\
$f_b $   & 0.09 \\
LOG $M_{\rmn{stellar}}$ & 9.91\\
\hline 
  \end{tabular}
  \label{tab:median_bestfit_bursts}
\end{table}

Our conclusion from this modelling work is that whilst regular SFHs
can provide an adequate explanation for approximately half of the
sample, they provide poor representations of the remaining galaxies.
By relaxing the assumption of a simple SFH, and instead allowing the
recent star-formation in the galaxies to evolve according to a simple
burst structure, these inconsistencies appear to be resolved.

\subsection{Comparisons with other samples}
\label{sec:compare-high-z}

The finding that the SFHs of a substantial fraction of the galaxies in
our dataset appear irregular has important implications for surveys
that measure SFRs in galaxies via UV measures, which we discuss in the
next section. Prior to that analysis, we first examine the results of
this survey with other redshift surveys as a consistency check to
confirm that the UV-selected galaxy properties make sense within the
broader galaxy population.

\begin{figure}
  \centering
  \includegraphics[width=80mm]{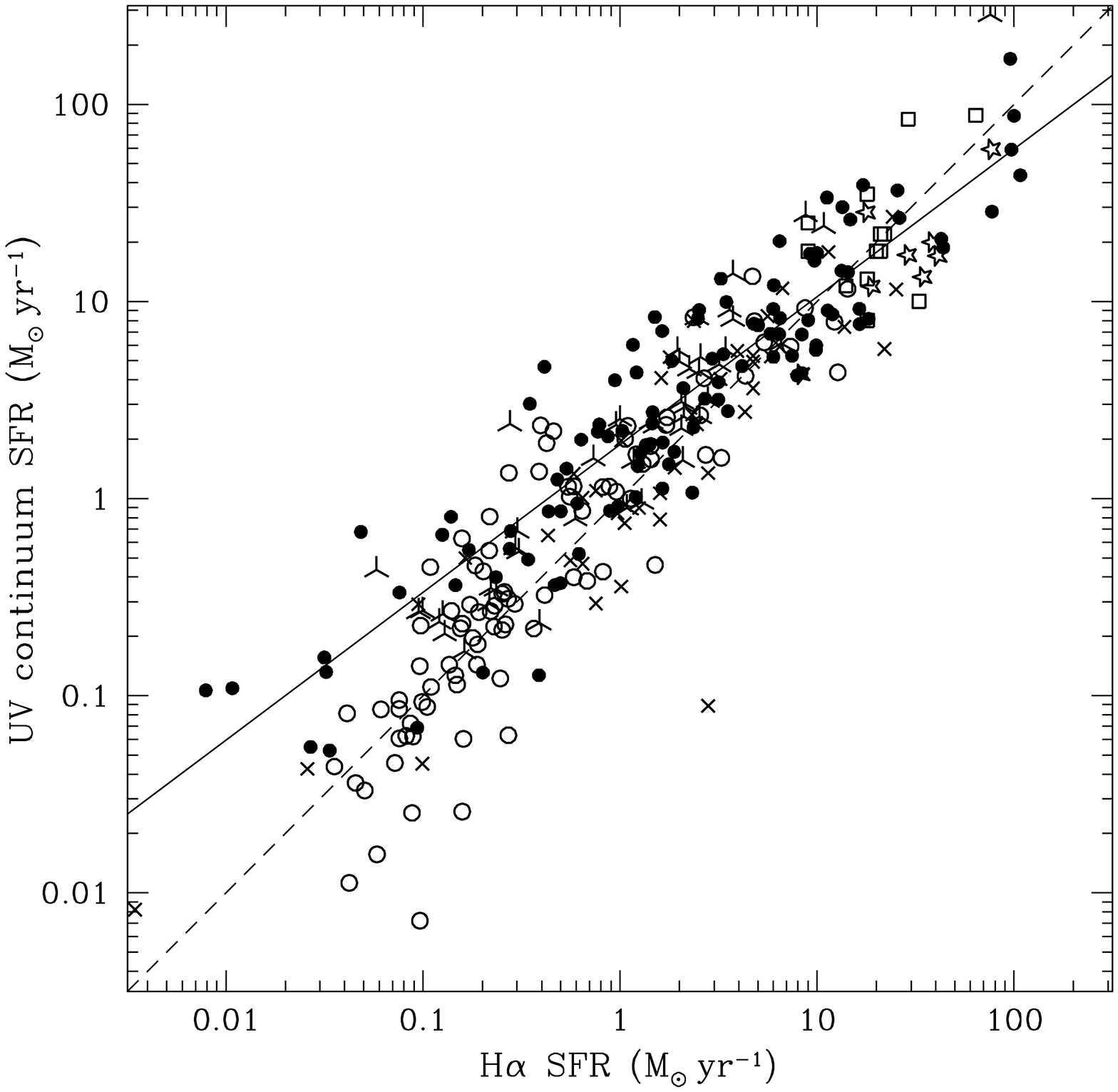}
  \caption{
    A comparison of the FOCA $0<z<0.4$ galaxies (data same as
    Fig.~\ref{fig:compare_sfrs}) with those from other samples at
    varying redshifts: Open circles: $z\simeq0.25$ and $z\simeq0.4$;
    \citet{2003A&A...402...65H}, Stars: $z\sim1$;
    \citet{1999MNRAS.306..843G}, Squares: $z\sim2.2$;
    \citet{2003ApJ...591..101E}, Square crosses: $z\sim0$;
    \citet{2001ApJ...548..681B}, Triangular crosses: $z\sim0$;
    \citet{2002A&A...383..801B}. The solid line shows the best-fit to
    the FOCA sample. All data are dust-corrected using the
    \citet{2000ApJ...533..682C} prescription.}
  \label{fig:compare_highz}
\end{figure}

Comparison of the results of this survey with other samples of
UV-selected galaxies at low-redshift are not yet possible due to the
lack of low-redshift UV observations (a situation soon to be rectified
via the \textsc{galex} experiment), and $U$-band observations must
suffice. However, at higher redshift a comparison is easier as
$U$-band fluxes are shifted into optical bandpasses. We compare here
with five different samples. The first, at $z\simeq0.25$ and
$z\simeq0.4$, is the emission-line (\ha) selected sample of
\citet{2003A&A...402...65H}. Next are two high-redshift samples of
\citet{1999MNRAS.306..843G} ($z\sim1$), who observe \ha\ fluxes for a
sample of $I$-band selected CFRS galaxies, and
\citet{2003ApJ...591..101E}, who select $z\sim2.2$ galaxies using a
photometric colour technique.  Finally, we show the two $z\simeq0$
samples of \citet{2001ApJ...548..681B} and
\citet{2002A&A...383..801B}. For all these samples, we take the
published \ha\ and UV-continuum fluxes, correct for dust using the
\citet{2000ApJ...533..682C} prescription if required, and convert to
SFRs using our cosmological model and the \pegase2\ conversions as
appropriate.  The samples are then plotted with the FOCA data in
Fig.~\ref{fig:compare_highz}.

In high-luminosity systems ($\rmn{SFR}\ga5\,\mperyr$), the only
systems probed by the high-redshift studies, the different samples
agree well, with some evidence for higher \ha-derived SFRs in these
systems. Whilst this could be caused by under-estimated dust
corrections, our alternative hypothesis of irregular SFHs can also
explain the observations: high \ha\ luminosity systems are likely near
a peak in SFR, where models predict that \ha\ and UV luminosities are
at least equal, or should even show an \ha\ excess (as \ha\ light
increases in a new starburst more rapidly than UV light).

Below SFRs of $\sim5\,\mperyr$, the properties of the different
samples begin to diverge. As noted in
Section~\ref{sec:comp-star-form}, the UV-selected sample shows a
general excess in the UV-derived SFRs. However, the two $z\simeq0$
samples demonstrate SFRs that agree well, while the \ha-selected
sample shows a larger scatter in the derived SFRs, with an \ha\ excess
at the faint end. Once again, a systematic under-estimation in the
dust extinction corrections can explain the \ha-excess at the faint
end, but this does not explain the UV-excess galaxies. Again, all of
the results can be consistently interpreted within a framework of
non-regular SFHs coupled with the various survey selection criteria:
\ha-selected galaxies are likely to be located at a phase in their SFH
where they are either near a peak of a starburst (in galaxies with
varying SFHs) and the \ha-derived SFR is greater than that from the
UV, or the galaxies will have regular SFHs and the \ha\ and UV SFRs
agree well. The $z\sim0$ galaxies, with a general optical selection
criteria, are likely to represent star-formation across normal Hubble
types and are less likely to possess the irregular SFHs found in the
FOCA sample which are selected by their UV light. We examine these
ideas further in Section~\ref{sec:impl-cosm-star}.

\subsection{Implications for measuring cosmic star-formation for flux-limited redshift surveys}
\label{sec:impl-cosm-star}

\begin{figure}
  \centering
   \includegraphics[width=85mm]{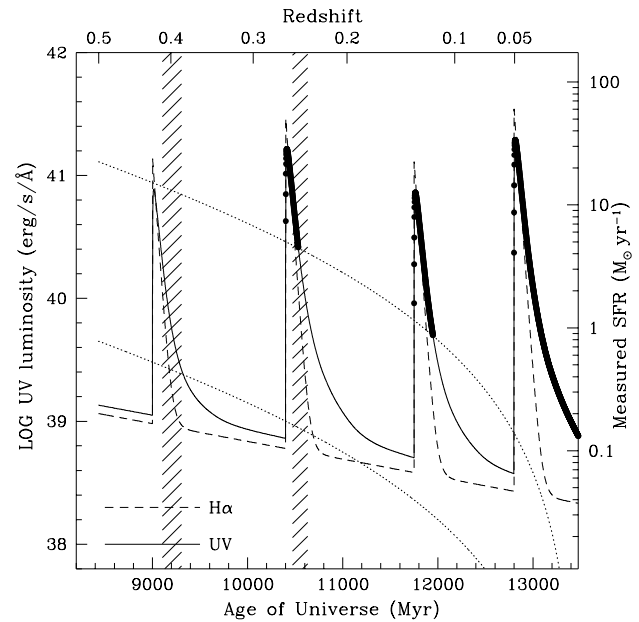}
  \caption{
    The visibility in the FOCA UV-selected and
    \citet{2003A&A...402...65H} emission-line selected surveys of a
    model galaxy undergoing a series of starbursts between $z=0.5$ and
    $z=0$. The continuous line denotes the UV luminosity; the dashed
    line the input SFR (the axes are scaled so that the UV luminosity
    and input SFR agree according to the conversion values in
    Table~\ref{tab:sfr_convert}). The filled circles show the periods
    during which a galaxy falls within the selection criteria of a
    FOCA-like survey, and the two overlaid dotted curves show the
    magnitude limit of the FOCA survey (upper curve) and the
    sensitivity limit of the \ha\ survey of
    \citet{2003A&A...402...65H} (lower curve). The vertical shaded
    areas at $\simeq0.25$ and $z\simeq0.4$ show the redshift
    sensitivity of \citet{2003A&A...402...65H}.  }
  \label{fig:visi_modelgalaxy_foca}
\end{figure}

The finding that a fraction of the galaxies in our sample do not
possess simple SFHs has implications for flux-limited redshift surveys
such as this. A galaxy that evolves with an intermittent SFH will
clearly brighten and dim over the course of its history. We
demonstrate this in Fig.~\ref{fig:visi_modelgalaxy_foca}, where we
show the UV evolution of a typical galaxy in our survey and the input
SFR(t) required to produce it, together with the flux limit of the
FOCA survey ($m_{\rmn{uv}}=20.75$ in the AB system). We also plot the
flux limit and survey parameters of the $z\simeq0.25$ and $z\simeq0.4$
\ha-selected survey of \citet{2003A&A...402...65H}.

The bursts of star-formation brighten the model galaxy above the UV
flux limit into the detectable magnitude range.  Without this boost, a
galaxy would otherwise not be detected in the UV unless located at a
low redshift. This bias in itself is not serious -- it is one of the
goals of this redshift survey to measure an integrated star-formation
density, including that fraction of star-formation that occurs in a
`burst mode' rather than a `continuous mode'. However, a more
important bias arises due to the time required for the UV light to die
away after a burst has completed, thus leading to higher measured SFRs
in the UV or $u$' than might be obtained via alternative diagnostic
measures, and higher measured SFRs than the `true' SFR in the galaxy.

We illustrate this using a model taking into account the flux limit of
the FOCA survey and \ha-selected survey.  We model a galaxy's SFH by
superimposing a series of bursts onto an exponentially declining SFH
between the ages in the galaxy's history that correspond to $z=0.5$
and $z=0$ in our cosmological model. The variations $\mathrm{UV}(t)$
and $\ha(t)$ are calculated, and using the mapping of
$\mathrm{time}\rightarrow z$, the apparent magnitude
$m_{\mathrm{uv}}(t)$ can then be computed from the luminosity distance
and an appropriate $k$-correction calculated from the synthetic
spectrum. From the flux limit of the two surveys, the visibility of
the galaxy in each survey can be found.

Fig.~\ref{fig:visi_modelgalaxy_foca} shows this model. There are
points over the model galaxy's evolution when, although the
instantaneous SFR of the galaxy lies below the flux limit of the UV
survey -- i.e. if this SFR were converted to a UV luminosity using a
simple constant-SFH assumption the galaxy would not be detected -- the
galaxy remains in the UV-selected survey due to the slower decline of
the UV light. As expected, at higher redshift a galaxy is
preferentially picked out if it is undergoing a burst of
star-formation and additionally if the galaxy is near the peak of the
particular starburst. At lower redshifts, we are able to view the
galaxy at later times into a particular starburst event, and only at
the lowest redshifts does the underlying (smoothly declining) galaxy
SED come into the FOCA survey.

\begin{figure}
\centering
\includegraphics[width=80mm]{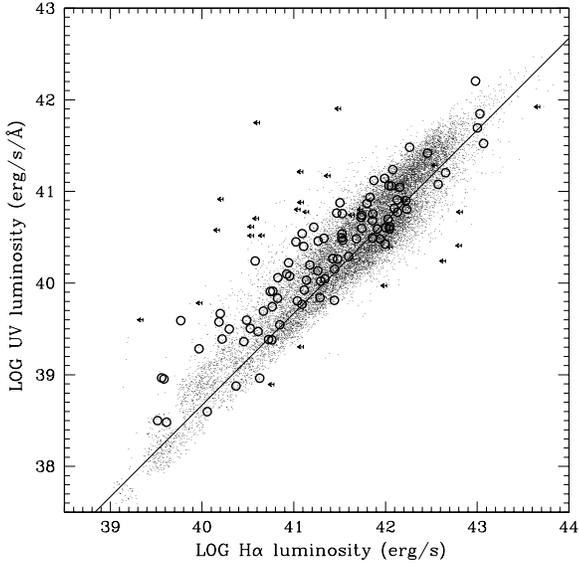}
\caption{
  An example simulation of the galaxies in our survey generated by
  superimposing starbursts of various masses and durations on a
  selection of underlying galaxy SFHs. By sampling a galaxy's history
  at random redshifts (weighted by the co-moving volume) during which
  the galaxy's apparent UV and \ha\ luminosities allow detection in
  the FOCA survey, distributions in the \lha\,--\,\luv\ plane (panel A
  from Fig.~\ref{fig:compare_sfrs}) can be generated and compared to
  the observations. Simulated galaxies are denoted by small dots,
  observed galaxies by open circles. The solid line denotes equality
  of SFRs from Table~\ref{tab:sfr_convert}. }
\label{fig:simulate_hauv}
\end{figure}

The effect of these forms of SFHs on the \ha--UV plane in a
UV-selected survey can be seen in Fig.~\ref{fig:simulate_hauv}. We run
200 simulations, generating SFHs containing a random number of bursts
(from zero to three) of varying mass and duration superimposed on
underlying histories with $\tau$ from 0.750 to 6\,Gyr between
$z=0-0.5$. The \ha\ and UV luminosities at 1\,Myr intervals are
calculated and given a random error distribution identical to that in
the survey. The galaxy is recorded (number-weighted by the co-moving
volume) if it meets the selection criteria of the UV survey, and then
example survey datasets drawn randomly from these results. The
distribution of 18000 simulated galaxies in \ha--UV space can then be
generated. An excellent match between the observed galaxy distribution
and the simulated galaxies can be seen.

Estimating how `biased' a determination of the SFR derived from a UV
continuum measurement in a particular galaxy is (i.e. how much this
measurement is over-estimated due to recent star-formation activity)
clearly requires a knowledge of the precise form and age of the
particular star-formation events. Though these parameters can, in
principle, be estimated using techniques such as those in
Section~\ref{sex:fitting-galaxy-seds}, this can only be done reliably
for a sub-sample of the galaxies -- and even then there is a
considerable uncertainty in the derived parameters due to the lack of
infra-red photometry (particularly the total stellar mass).

Instead, we can estimate the effect of the recent SFH in a purely
statistical manner based on the entire sample. We can use the plots of
Fig.~\ref{fig:compare_sfrs} and the statistical tests performed on
them to derive relations between UV or $u$' luminosities and the
\textit{instantaneous} (\ha-derived) SFR. By fitting the data across
the range of SFRs probed in this study, we find:

\begin{eqnarray}
\label{eq:2}
\log(\luv)&=&40.03+0.75\log(\rmn{SFR}_{\ha})\nonumber\\
\rightarrow \rmn{SFR}&=&(\frac{\luv}{1.07\times10^{40}})^{1.33}\,\,\mperyr
\end{eqnarray}

\noindent
for the UV luminosities, and

\begin{eqnarray}
\label{eq:3}
\log(\lu)&=&39.53+0.74\log(\rmn{SFR}_{\ha})\nonumber\\
\rightarrow \rmn{SFR}&=&(\frac{\luv}{3.39\times10^{39}})^{1.35}\,\,\mperyr
\end{eqnarray}

\noindent
for the $u$' luminosities, where the luminosities are dust-corrected
and measured in \ergsa. This correction is valid over SFRs of
approximately 0.1 to 100\,\mperyr\ for galaxies in this survey
selected at 2000\,\AA.  Whilst similar relational forms likely hold
for other redshift surveys, the exact form will depend on the
selection criteria and hence the number of galaxies with `normal' SFHs
admitted into the survey. Even for other UV-selected surveys,
selection at a shorter wavelength will be less affected by the effects
of starbursts as the UV light dies away more quickly, whilst longer
wavelengths could suffer considerable bias.

Understanding the precise impact of this bias on analyses such as the
Madau plot is a complex problem. The corrections given above are only
applicable to individual galaxies, not to the integrated luminosity
density of the survey as a whole, as the standard luminosity function
\citet{1976ApJ...203..297S} parameters ($\alpha$,
$M_{\ast}$/$L_{\ast}$ and $\phi_{\ast}$) will be affected in different
ways by correcting for this effect. This correction tends to make the
SFR in fainter galaxies lower, and hence the appropriate luminosity
function slope will be flatter, or $\alpha$ will become less negative,
likely leading to a slight decrease in the calculated integrated
star-formation density.  The luminosity functions and light densities
of this sample will be addressed in a forthcoming paper (Treyer,
Sullivan \& Ellis, in preparation).

The conclusion from this study is that integrated luminosity densities
are a poor guide to the complex physical processes at play in these
systems. Star-formation densities derived from these measures need to
be carefully calibrated, and the effects of different biases at work
more fully understood, before the results of flux-limited surveys can
be fairly compared.

\section{Conclusions}
\label{sec:conclusions}

In this paper, we have presented the results of panoramic wide-field
optical imaging of a UV-selected galaxy redshift survey, with the aim
of further investigating the nature of star-formation in local $z<0.4$
star-forming and starburst galaxies. We have found the following:

\begin{enumerate}
  
\item New $u'Br'$ photometry, supplemented by DPOSS data, have allowed
  us to derive aperture corrections for our \ha\ fluxes, as well as
  more reliable SEDs for all the sample galaxies. There remains a
  small fraction ($\le 8$ per cent) of galaxies exhibiting $UV-B$
  colours bluer than most starburst models, whilst possessing colours
  typical of standard SEDs at longer wavelength.
  
\item We have investigated the dependence of the \ha, UV and $u$'
  luminosity to SFR conversion factors on the IMF and mass ranges, the
  stellar metallicity and the time since onset of star formation,
  assuming constant SFHs.  Though varying these parameters can have a
  large effect on individual conversion values, the effect is smaller
  when considering ratios of these values (e.g. UV/\ha).
  
\item Taking advantage of our new dataset, we update the study of
  \citet{2000MNRAS.312..442S} and compare SFRs derived from UV or $u$'
  and \ha\ luminosities. Assuming simple SFHs, we show the scatter and
  non-unity best-fitting slopes observed are unlikely to be primarily
  generated by varying the above parameters.
  
\item We show that models including a burst or increased period of
  star-formation superimposed on an otherwise smooth underlying SFH
  provide much better fits to the data set of $\sim$50\% of the
  galaxies (\ha, UV and $u$' luminosities, Balmer Break, \ha\ EW and
  colours) than a smooth SFH alone.
 
\item Such burst modes of star formation lead to an overestimate of
  SFRs derived from UV luminosities in lower luminosity systems, as UV
  light from less massive stars will still be present after a burst
  has died away. We propose a simple statistical correction for
  UV-selected surveys based on the `true' SFRs as derived from \ha\ 
  luminosities which originate only from the most massive stars.

\end{enumerate}

\section*{Acknowledgements}

MS acknowledges support from a PPARC fellowship. We thank Roy Gal for
providing us with the DPOSS data for the SA57 and A1367 survey fields,
Andrew Firth for assisting with the CFHT data collection, Andrew
Hopkins for useful discussions, and St\'ephane Arnouts for his advice
on using \sex. We are grateful to Yannick Mellier and Mireille Dantel
for their invaluable assistance with the CFHT data reduction at the
TERAPIX centre.  We thank Mark Metzger and Rob Simcoe for assistance
using the LFC on the Palomar 200-in telescope. We also thank Chuck
Steidel and Jean-Charles Cuillandre for providing Palomar and CFH12k
filter response curves respectively.  The WIYN Observatory is a joint
facility of the University of Wisconsin-Madison, Indiana University,
Yale University, and the National Optical Astronomy Observatories. The
William Herschel Telescope is operated on the island of La Palma by
the Isaac Newton Group in the Spanish Observatorio del Roque de los
Muchachos of the Instituto de Astrofisica de Canarias. The
Canada-France-Hawaii Telescope (CFHT) is operated by the National
Research Council of Canada, the Institut National des Science de
l'Univers of the Centre National de la Recherche Scientifique of
France, and the University of Hawaii.

\label{lastpage}

\begin{thebibliography}{71}
\expandafter\ifx\csname natexlab\endcsname\relax\def\natexlab#1{#1}\fi

\bibitem[{{Afonso} {et~al.}(2003){Afonso}, {Hopkins}, {Mobasher}, \&
  {Almeida}}]{2003astroph0307175A}
{Afonso}, J., {Hopkins}, A., {Mobasher}, B., \& {Almeida}, C. 2003, in press,
  ApJ, astro--ph/0307175

\bibitem[{{Bell}(2002)}]{2002ApJ...577..150B}
{Bell}, E.~F. 2002, \apj, 577, 150

\bibitem[{{Bell} \& {Kennicutt}(2001)}]{2001ApJ...548..681B}
{Bell}, E.~F. \& {Kennicutt}, R.~C. 2001, \apj, 548, 681

\bibitem[{{Bertin} \& {Arnouts}(1996)}]{1996A&AS..117..393B}
{Bertin}, E. \& {Arnouts}, S. 1996, \aaps, 117, 393

\bibitem[{{Blain} {et~al.}(1999){Blain}, {Smail}, {Ivison}, \&
  {Kneib}}]{1999MNRAS.302..632B}
{Blain}, A.~W., {Smail}, I., {Ivison}, R.~J., \& {Kneib}, J. 1999, \mnras, 302,
  632

\bibitem[{{Brown} {et~al.}(2000){Brown}, {Kenyon}, {Geller}, \&
  {Fabricant}}]{2000ApJ...540L..83B}
{Brown}, W.~R., {Kenyon}, S.~J., {Geller}, M.~J., \& {Fabricant}, D.~G. 2000,
  \apjl, 540, L83

\bibitem[{{Bruzual}(1983)}]{1983ApJ...273..105B}
{Bruzual}, A.~G. 1983, \apj, 273, 105

\bibitem[{{Buat} {et~al.}(2002){Buat}, {Boselli}, {Gavazzi}, \&
  {Bonfanti}}]{2002A&A...383..801B}
{Buat}, V., {Boselli}, A., {Gavazzi}, G., \& {Bonfanti}, C. 2002, \aap, 383,
  801

\bibitem[{{Calzetti} {et~al.}(2000){Calzetti}, {Armus}, {Bohlin}, {Kinney},
  {Koornneef}, \& {Storchi-Bergmann}}]{2000ApJ...533..682C}
{Calzetti}, D., {Armus}, L., {Bohlin}, R.~C., {Kinney}, A.~L., {Koornneef}, J.,
  \& {Storchi-Bergmann}, T. 2000, \apj, 533, 682

\bibitem[{{Calzetti} {et~al.}(1994){Calzetti}, {Kinney}, \&
  {Storchi-Bergmann}}]{1994ApJ...429..582C}
{Calzetti}, D., {Kinney}, A.~L., \& {Storchi-Bergmann}, T. 1994, \apj, 429, 582

\bibitem[{{Cardelli} {et~al.}(1989){Cardelli}, {Clayton}, \&
  {Mathis}}]{1989ApJ...345..245C}
{Cardelli}, J.~A., {Clayton}, G.~C., \& {Mathis}, J.~S. 1989, \apj, 345, 245

\bibitem[{{Charlot} \& {Longhetti}(2001)}]{2001MNRAS.323..887C}
{Charlot}, S.~. \& {Longhetti}, M. 2001, \mnras, 323, 887

\bibitem[{{Coleman} {et~al.}(1980){Coleman}, {Wu}, \&
  {Weedman}}]{1980ApJS...43..393C}
{Coleman}, G.~D., {Wu}, C.~., \& {Weedman}, D.~W. 1980, \apjs, 43, 393

\bibitem[{{Condon}(1992)}]{1992ARA&A..30..575C}
{Condon}, J.~J. 1992, \araa, 30, 575

\bibitem[{{Connolly} {et~al.}(1995){Connolly}, {Szalay}, {Bershady}, {Kinney},
  \& {Calzetti}}]{1995AJ....110.1071C}
{Connolly}, A.~J., {Szalay}, A.~S., {Bershady}, M.~A., {Kinney}, A.~L., \&
  {Calzetti}, D. 1995, \aj, 110, 1071

\bibitem[{{Connolly} {et~al.}(1997){Connolly}, {Szalay}, {Dickinson},
  {Subbarao}, \& {Brunner}}]{1997ApJ...486L..11C}
{Connolly}, A.~J., {Szalay}, A.~S., {Dickinson}, M., {Subbarao}, M.~U., \&
  {Brunner}, R.~J. 1997, \apjl, 486, L11

\bibitem[{{Contini} {et~al.}(2002){Contini}, {Treyer}, {Sullivan}, \&
  {Ellis}}]{2002MNRAS.330...75C}
{Contini}, T., {Treyer}, M.~A., {Sullivan}, M., \& {Ellis}, R.~S. 2002, \mnras,
  330, 75

\bibitem[{{Cowie} {et~al.}(1999){Cowie}, {Songaila}, \&
  {Barger}}]{1999AJ....118..603C}
{Cowie}, L.~L., {Songaila}, A., \& {Barger}, A.~J. 1999, \aj, 118, 603

\bibitem[{{Cram} {et~al.}(1998){Cram}, {Hopkins}, {Mobasher}, \&
  {Rowan-Robinson}}]{1998ApJ...507..155C}
{Cram}, L., {Hopkins}, A., {Mobasher}, B., \& {Rowan-Robinson}, M. 1998, \apj,
  507, 155

\bibitem[{{Cuillandre} {et~al.}(2000){Cuillandre}, {Luppino}, {Starr}, \&
  {Isani}}]{2000SPIE.4008.1010C}
{Cuillandre}, J., {Luppino}, G.~A., {Starr}, B.~M., \& {Isani}, S. 2000,
  \procspie, 4008, 1010

\bibitem[{{Donas} {et~al.}(1987){Donas}, {Deharveng}, {Laget}, {Milliard}, \&
  {Huguenin}}]{1987A&A...180...12D}
{Donas}, J., {Deharveng}, J.~M., {Laget}, M., {Milliard}, B., \& {Huguenin}, D.
  1987, \aap, 180, 12

\bibitem[{{Erb} {et~al.}(2003){Erb}, {Shapley}, {Steidel}, {Pettini},
  {Adelberger}, {Hunt}, {Moorwood}, \& {Cuby}}]{2003ApJ...591..101E}
{Erb}, D.~K., {Shapley}, A.~E., {Steidel}, C.~C., {Pettini}, M., {Adelberger},
  K.~L., {Hunt}, M.~P., {Moorwood}, A.~F.~M., \& {Cuby}, J. 2003, \apj, 591,
  101

\bibitem[{{Fanelli} {et~al.}(1988){Fanelli}, {O'Connell}, \&
  {Thuan}}]{1988ApJ...334..665F}
{Fanelli}, M.~N., {O'Connell}, R.~W., \& {Thuan}, T.~X. 1988, \apj, 334, 665

\bibitem[{{Fioc} \& {Rocca-Volmerange}(1997)}]{1997A&A...326..950F}
{Fioc}, M. \& {Rocca-Volmerange}, B. 1997, \aap, 326, 950

\bibitem[{{Fioc} \& {Rocca-Volmerange}(1999)}]{1999astro.ph.9912179}
{Fioc}, M. \& {Rocca-Volmerange}, B. 1999, in astro-ph, astro--ph/9912179

\bibitem[{{Folkes} {et~al.}(1999){Folkes}, {Ronen}, {Price}, {Lahav},
  {Colless}, {Maddox}, {Deeley}, {Glazebrook}, {Bland-Hawthorn}, {Cannon},
  {Cole}, {Collins}, {Couch}, {Driver}, {Dalton}, {Efstathiou}, {Ellis},
  {Frenk}, {Kaiser}, {Lewis}, {Lumsden}, {Peacock}, {Peterson}, {Sutherland},
  \& {Taylor}}]{1999MNRAS.308..459F}
{Folkes}, S., et al. 1999, \mnras, 308, 459

\bibitem[{{Fukugita} {et~al.}(1995){Fukugita}, {Shimasaku}, \&
  {Ichikawa}}]{1995PASP..107..945F}
{Fukugita}, M., {Shimasaku}, K., \& {Ichikawa}, T. 1995, \pasp, 107, 945

\bibitem[{{Gal} {et~al.}(????){Gal}, {de Carvalho}, {Odewahn}, {Djorgovski},
  {Mahabal}, {Brunner}, \& {Lopes}}]{2002astro.ph.10298G}
{Gal}, R.~R., {de Carvalho}, R.~R., {Odewahn}, S.~C., {Djorgovski}, S.~G.,
  {Mahabal}, A.~A., {Brunner}, R.~J., \& {Lopes}, P.~A.~A. ????, in press, AJ

\bibitem[{{Gallagher} {et~al.}(1989){Gallagher}, {Hunter}, \&
  {Bushouse}}]{1989AJ.....97..700G}
{Gallagher}, J.~S., {Hunter}, D.~A., \& {Bushouse}, H. 1989, \aj, 97, 700

\bibitem[{{Gallego} {et~al.}(1995){Gallego}, {Zamorano}, {Aragon-Salamanca}, \&
  {Rego}}]{1995ApJ...455L...1G}
{Gallego}, J., {Zamorano}, J., {Aragon-Salamanca}, A., \& {Rego}, M. 1995,
  \apjl, 455, L1

\bibitem[{{Glazebrook} {et~al.}(1999){Glazebrook}, {Blake}, {Economou},
  {Lilly}, \& {Colless}}]{1999MNRAS.306..843G}
{Glazebrook}, K., {Blake}, C., {Economou}, F., {Lilly}, S., \& {Colless}, M.
  1999, \mnras, 306, 843

\bibitem[{{Haarsma} {et~al.}(2000){Haarsma}, {Partridge}, {Windhorst}, \&
  {Richards}}]{2000ApJ...544..641H}
{Haarsma}, D.~B., {Partridge}, R.~B., {Windhorst}, R.~A., \& {Richards}, E.~A.
  2000, \apj, 544, 641

\bibitem[{{Heyl} {et~al.}(1997){Heyl}, {Colless}, {Ellis}, \&
  {Broadhurst}}]{1997MNRAS.285..613H}
{Heyl}, J., {Colless}, M., {Ellis}, R.~S., \& {Broadhurst}, T. 1997, \mnras,
  285, 613

\bibitem[{{Hippelein} {et~al.}(2003){Hippelein}, {Maier}, {Meisenheimer},
  {Wolf}, {Fried}, {von Kuhlmann}, {K{\" u}mmel}, {Phleps}, \& {R{\"
  o}ser}}]{2003A&A...402...65H}
{Hippelein}, H., {Maier}, C., {Meisenheimer}, K., {Wolf}, C., {Fried}, J.~W.,
  {von Kuhlmann}, B., {K{\" u}mmel}, M., {Phleps}, S., \& {R{\" o}ser}, H.-J.
  2003, \aap, 402, 65

\bibitem[{{Hopkins} {et~al.}(2001){Hopkins}, {Connolly}, {Haarsma}, \&
  {Cram}}]{2001AJ....122..288H}
{Hopkins}, A.~M., {Connolly}, A.~J., {Haarsma}, D.~B., \& {Cram}, L.~E. 2001,
  \aj, 122, 288

\bibitem[{{Johnson} \& {Morgan}(1953)}]{1953ApJ...117..313J}
{Johnson}, H.~L. \& {Morgan}, W.~W. 1953, \apj, 117, 313

\bibitem[{{Kennicutt}(1983)}]{1983ApJ...272...54K}
{Kennicutt}, R.~C. 1983, \apj, 272, 54

\bibitem[{{Kennicutt}(1998)}]{1998ARA&A..36..189K}
---. 1998, \araa, 36, 189

\bibitem[{{Kewley} {et~al.}(2002){Kewley}, {Geller}, {Jansen}, \&
  {Dopita}}]{2002AJ....124.3135K}
{Kewley}, L.~J., {Geller}, M.~J., {Jansen}, R.~A., \& {Dopita}, M.~A. 2002,
  \aj, 124, 3135

\bibitem[{{Kron}(1980)}]{1980ApJS...43..305K}
{Kron}, R.~G. 1980, \apjs, 43, 305

\bibitem[{{Kroupa}(2001)}]{2001MNRAS.322..231K}
{Kroupa}, P. 2001, \mnras, 322, 231

\bibitem[{{Landolt}(1992)}]{1992AJ....104..340L}
{Landolt}, A.~U. 1992, \aj, 104, 340

\bibitem[{{Lilly} {et~al.}(1996){Lilly}, {Le Fevre}, {Hammer}, \&
  {Crampton}}]{1996ApJ...460L...1L}
{Lilly}, S.~J., {Le Fevre}, O., {Hammer}, F., \& {Crampton}, D. 1996, \apjl,
  460, L1

\bibitem[{{Madgwick} {et~al.}(2002){Madgwick}, {Lahav}, {Baldry}, {Baugh},
  {Bland-Hawthorn}, {Bridges}, {Cannon}, {Cole}, {Colless}, {Collins}, {Couch},
  {Dalton}, {De Propris}, {Driver}, {Efstathiou}, {Ellis}, {Frenk},
  {Glazebrook}, {Jackson}, {Lewis}, {Lumsden}, {Maddox}, {Norberg}, {Peacock},
  {Peterson}, {Sutherland}, \& {Taylor}}]{2002MNRAS.333..133M}
{Madgwick}, D.~S., et al. 2002,
  \mnras, 333, 133

\bibitem[{{Mas-Hesse} \& {Kunth}(1999)}]{1999A&A...349..765M}
{Mas-Hesse}, J.~M. \& {Kunth}, D. 1999, \aap, 349, 765

\bibitem[{{Massey} \& {Gronwall}(1990)}]{1990ApJ...358..344M}
{Massey}, P. \& {Gronwall}, C. 1990, \apj, 358, 344

\bibitem[{{Massey} {et~al.}(1988){Massey}, {Strobel}, {Barnes}, \&
  {Anderson}}]{1988ApJ...328..315M}
{Massey}, P., {Strobel}, K., {Barnes}, J.~V., \& {Anderson}, E. 1988, \apj,
  328, 315

\bibitem[{{Milliard} {et~al.}(1992){Milliard}, {Donas}, {Laget}, {Armand}, \&
  {Vuillemin}}]{1992A&A...257...24M}
{Milliard}, B., {Donas}, J., {Laget}, M., {Armand}, C., \& {Vuillemin}, A.
  1992, \aap, 257, 24

\bibitem[{{Mink}(1999)}]{1999adass...8..498M}
{Mink}, D.~J. 1999, in ASP Conf. Ser. 172: Astronomical Data Analysis Software
  and Systems VIII, Vol.~8, 498

\bibitem[{{Oke}(1974)}]{1974ApJS...27...21O}
{Oke}, J.~B. 1974, \apjs, 27, 21

\bibitem[{{Oke} \& {Gunn}(1983)}]{1983ApJ...266..713O}
{Oke}, J.~B. \& {Gunn}, J.~E. 1983, \apj, 266, 713

\bibitem[{{Poggianti}(1997)}]{1997A&AS..122..399P}
{Poggianti}, B.~M. 1997, \aaps, 122, 399

\bibitem[{{Rowan-Robinson} {et~al.}(1997){Rowan-Robinson}, {Mann}, {Oliver},
  {Efstathiou}, {Eaton}, {Goldschmidt}, {Mobasher}, {Serjeant}, {Sumner},
  {Danese}, {Elbaz}, {Franceschini}, {Egami}, {Kontizas}, {Lawrence},
  {McMahon}, {Norgaard-Nielsen}, {Perez-Fournon}, \&
  {Gonzalez-Serrano}}]{1997MNRAS.289..490R}
{Rowan-Robinson}, M., et al. 1997, \mnras, 289, 490

\bibitem[{{Salpeter}(1955)}]{1955ApJ...121..161S}
{Salpeter}, E.~E. 1955, \apj, 121, 161

\bibitem[{{Scalo}(1998)}]{1998simf.conf..201S}
{Scalo}, J. 1998, in ASP Conf. Ser. 142: The Stellar Initial Mass Function, ed.
  G. Gilmore \& D.Howell (San Francisco: ASP), 201

\bibitem[{{Schechter}(1976)}]{1976ApJ...203..297S}
{Schechter}, P. 1976, \apj, 203, 297

\bibitem[{{Schlegel} {et~al.}(1998){Schlegel}, {Finkbeiner}, \&
  {Davis}}]{1998ApJ...500..525S}
{Schlegel}, D.~J., {Finkbeiner}, D.~P., \& {Davis}, M. 1998, \apj, 500, 525

\bibitem[{{Searle} {et~al.}(1973){Searle}, {Sargent}, \&
  {Bagnuolo}}]{1973ApJ...179..427S}
{Searle}, L., {Sargent}, W.~L.~W., \& {Bagnuolo}, W.~G. 1973, \apj, 179, 427

\bibitem[{{Simcoe} {et~al.}(2000){Simcoe}, {Metzger}, {Small}, \&
  {Araya}}]{2000AAS...196.5209S}
{Simcoe}, R.~A., {Metzger}, M.~R., {Small}, T.~A., \& {Araya}, G. 2000, in
  American Astronomical Society Meeting, Vol. 196, 5209

\bibitem[{{Steidel} \& {Hamilton}(1993)}]{1993AJ....105.2017S}
{Steidel}, C.~C. \& {Hamilton}, D. 1993, \aj, 105, 2017

\bibitem[{{Stone} {et~al.}(1999){Stone}, {Pier}, \&
  {Monet}}]{1999AJ....118.2488S}
{Stone}, R.~C., {Pier}, J.~R., \& {Monet}, D.~G. 1999, \aj, 118, 2488

\bibitem[{{Sullivan}(2002)}]{2002Obs...122..307S}
{Sullivan}, M. 2002, Ph.D. Thesis (Univeristy of Cambridge); abstract appears
  in The Observatory, 122, 307

\bibitem[{{Sullivan} {et~al.}(2001){Sullivan}, {Mobasher}, {Chan}, {Cram},
  {Ellis}, {Treyer}, \& {Hopkins}}]{2001ApJ...558...72S}
{Sullivan}, M., {Mobasher}, B., {Chan}, B., {Cram}, L., {Ellis}, R., {Treyer},
  M., \& {Hopkins}, A. 2001, \apj, 558, 72

\bibitem[{{Sullivan} {et~al.}(2000){Sullivan}, {Treyer}, {Ellis}, {Bridges},
  {Milliard}, \& {Donas}}]{2000MNRAS.312..442S}
{Sullivan}, M., {Treyer}, M.~A., {Ellis}, R.~S., {Bridges}, T.~J., {Milliard},
  B., \& {Donas}, J.~. 2000, \mnras, 312, 442

\bibitem[{{Tresse} \& {Maddox}(1998)}]{1998ApJ...495..691T}
{Tresse}, L. \& {Maddox}, S.~J. 1998, \apj, 495, 691

\bibitem[{{Tresse} {et~al.}(2002){Tresse}, {Maddox}, {Le F{\` e}vre}, \&
  {Cuby}}]{2002MNRAS.337..369T}
{Tresse}, L., {Maddox}, S.~J., {Le F{\` e}vre}, O., \& {Cuby}, J.-G. 2002,
  \mnras, 337, 369

\bibitem[{{Treyer} {et~al.}(1998){Treyer}, {Ellis}, {Milliard}, {Donas}, \&
  {Bridges}}]{1998MNRAS.300..303T}
{Treyer}, M.~A., {Ellis}, R.~S., {Milliard}, B., {Donas}, J., \& {Bridges},
  T.~J. 1998, \mnras, 300, 303

\bibitem[{{Valdes}(1998)}]{1998adass...7...53V}
{Valdes}, F.~G. 1998, in ASP Conf. Ser. 145: Astronomical Data Analysis
  Software and Systems VII, Vol.~7, 53

\bibitem[{{Wilson} {et~al.}(2002){Wilson}, {Cowie}, {Barger}, \&
  {Burke}}]{2002AJ....124.1258W}
{Wilson}, G., {Cowie}, L.~L., {Barger}, A.~J., \& {Burke}, D.~J. 2002, \aj,
  124, 1258

\bibitem[{{Yoshii} \& {Takahara}(1988)}]{1988ApJ...326....1Y}
{Yoshii}, Y. \& {Takahara}, F. 1988, \apj, 326, 1

\bibitem[{{Zaritsky} {et~al.}(1995){Zaritsky}, {Zabludoff}, \&
  {Willick}}]{1995AJ....110.1602Z}
{Zaritsky}, D., {Zabludoff}, A.~I., \& {Willick}, J.~A. 1995, \aj, 110, 1602

\end{thebibliography}
\end{document}